\documentclass[aps,prx,twocolumn,showpacs,superscriptaddress, longbibliography]{revtex4-1}
\usepackage{times}
\usepackage{amsmath}
\usepackage{amsfonts}
\usepackage{amssymb}
\usepackage{dsfont}
\usepackage{graphicx}

\usepackage{epsfig}
\usepackage{verbatim}
\usepackage{bm}
\usepackage{mathrsfs}
\usepackage{yfonts}

\usepackage{pst-all}
\usepackage{leftidx}

\newcommand{\mb}[1]{\mathbf{#1}}
\newcommand{\mr}[1]{\mathrm{#1}}

\def\U{\mathrm{U}(1)}
\def\H{\mathcal{H}}
\def\Z{\mathbb{Z}}

\newcommand{\coho}[1]{\textswab{#1}}
\newcommand{\cohosub}[1]{\scalebox{0.7}{\textswab{#1}}}

\begin{document}

\title{Time-reversal and spatial reflection symmetry localization anomalies \\ in (2+1)D topological phases of matter}
\author{Maissam Barkeshli}
\affiliation{Department of Physics, Condensed Matter Theory Center, University of Maryland, College Park, Maryland 20742, USA}
\affiliation{Joint Quantum Institute, University of Maryland, College Park, Maryland 20742, USA}
\author{Meng Cheng}
\affiliation{Department of Physics, Yale University, New Haven, CT 06511-8499, USA}

\begin{abstract}
We study a class of anomalies associated with time-reversal and spatial reflection symmetry in (2+1)D bosonic topological
phases of matter. In these systems, the topological quantum numbers of the quasiparticles, such as the fusion 
rules and braiding statistics, possess a $\mathbb{Z}_2$ symmetry which can be associated with 
either time-reversal (denoted $\mathbb{Z}_2^{\bf T})$ or spatial reflections.
Under this symmetry, correlation functions of all Wilson loop operators in the low energy topological quantum field theory (TQFT)
are invariant. However, the theories that we study possess a severe anomaly associated with the failure 
to consistently localize the symmetry action to the quasiparticles, precluding even defining a consistent
notion of symmetry fractionalization in such systems. We present simple sufficient conditions 
which determine when $\mathbb{Z}_2^{\bf T}$ symmetry localization anomalies exist in general. 
We present an infinite series of TQFTs with such anomalies, some examples of which include 
$\mr{USp}(4)_2$ Chern-Simons (CS) theory and $\mr{SO(4)}_4$ CS theory. The theories that we find with these 
$\mathbb{Z}_2^{\bf T}$ anomalies can all be obtained by gauging the unitary $\mathbb{Z}_2$ subgroup 
of a different TQFT with a $\mathbb{Z}_4^{\bf T}$ symmetry. We further show that 
the anomaly can be resolved in several distinct ways: (1) the true symmetry of the theory is $\mathbb{Z}_4^{\bf T}$,
or (2) the theory can be considered to be a theory of fermions, with ${\bf T}^2 = (-1)^{N_f}$ corresponding 
to fermion parity. Finally, we demonstrate that theories with the $\mathbb{Z}_2^{\bf T}$ localization anomaly can be compatible with 
$\mathbb{Z}_2^{\bf T}$ if they are  ``pseudo-realized'' at the surface of a (3+1)D symmetry-enriched topological 
phase. The ``pseudo-realization'' refers to the fact that the bulk (3+1)D system is described by a dynamical 
$\mathbb{Z}_2$ gauge theory and thus only a subset of the quasi-particles are truly confined to the surface. 
\end{abstract}

\maketitle

\section{Introduction}

There has recently been immense progress in understanding the interplay of global symmetries and topological degrees
of freedom in physics. From the perspective of condensed matter physics, this has led to advances in our understanding
of the distinct possible gapped quantum phases of matter by providing the theoretical framework for describing their 
universal long-wavelength properties and leading to a host of topological invariants that can distinguish such phases. 
On the other hand, many of these developments can be viewed entirely within the framework of quantum field theory, 
and have led to advances in our understanding of global symmetries in topological quantum field theory. 

In two and higher spatial dimensions, the study of topological phases of matter with global symmetries is still in progress. 
Even without any global symmetry, gapped quantum systems can still form distinct phases of matter, characterized
by their topological order. These states are distinguished by various exotic properties, including topologically non-trivial
quasiparticle excitations with fractional or non-Abelian braiding statistics, robust topological ground state 
degeneracies, and protected gapless edge modes.\cite{wen04,nayak2008,wang2008}

The intrinsic topological order in (2+1)D states is believed to be fully characterized by two objects: (1) the chiral central
charge $c_-$ of the phase, which describes the chiral energy transport along the (1+1)D boundary of the system, 
and (2) an algebraic theory $\mathcal{C}$, known as a unitary modular tensor category (UMTC),\cite{moore1989b,witten1989} which encapsulates the topological 
properties of the quasiparticles, such as their topological spins, fusion rules, and braiding transformations.

In the presence of a global symmetry group $G$, it is important to distinguish two types of phases: (1) invertible~\cite{freed2014, freed2016}, or 
short-range entangled states,\cite{ChenPRB2010} and (2) long-range entangled, topologically ordered states. Invertible states have
the property that given the state, there is an ``inverse'' state which, when the two are combined together, can be
transformed into a trivial product state by a finite-depth (in the limit of infinite system size) local unitary quantum
circuit (or, equivalently, by adiabatically tuning the parameters of the Hamiltonian without closing the bulk energy gap). 
In (2+1)D, these correspond to cases where the UMTC $\mathcal{C}$ is trivial. A special class of invertible states 
are symmetry-protected topological (SPT) states~\cite{XieScience2012, chen2013, lu2012, kapustin2014, senthil2015}. SPT 
states have the property that the state can be transformed 
into a product state by a finite-depth local unitary quantum circuit that breaks the $G$ symmetry~\cite{ChenPRB2010}; a non-trivial 
SPT state cannot be transformed into a product state by a $G$-symmetric finite-depth local unitary quantum circuit. 

Long-range entangled, or topologically ordered, states cannot be transformed into a product state by any finite-depth
local unitary quantum circuit, even in the absence of any global symmetry. In the presence of a global symmetry group
$G$, the class of topologically ordered states is refined into symmetry-enriched topological (SET) 
states~\cite{wen2002psg, wen04, LevinPRB2012, essin2013, mesaros2013, lu2013, barkeshli2014SDG, Tarantino_SET, Chen2014,teo2015, Lan2015, Lan2016}. Different
SETs with the same intrinsic topological order differ in the way the global symmetry interplays with the topological order.
This leads to different ways that the topologically non-trivial quasiparticles can carry fractional quantum numbers of the 
symmetry group~\cite{wen2002psg,essin2013, barkeshli2014SDG, Chen2014}, and different topological properties of symmetry defects~\cite{barkeshli2014SDG, Chen2014, Tarantino_SET, teo2015}. 

In Ref. \onlinecite{barkeshli2014SDG}, a systematic theoretical framework for characterizing symmetry-enriched
topological phases was presented. Each (2+1)D topological phase has a group of symmetries (possibly emergent), denoted 
$\text{Aut}(\mathcal{C})$, and which we refer to as the group of \it topological symmetries\rm. This is the
group of symmetries of the long wavelength effective topological quantum field theory (TQFT). It consists of permutations of the anyon types
which keeps their topological spins, fusion rules, and braiding statistics invariant (up to certain complex 
conjugations that are required for space-time parity reversing symmetries). Even in the absence of a global 
symmetry $G$, a topological phase of matter can have a non-trivial Aut$(\mathcal{C})$, which describes the 
group of \it emergent \rm symmetries of the topological quantum numbers of long wavelength degrees of 
freedom in the system. For example, for the $1/m$ Laughlin fractional quantum Hall (FQH) states, this includes
the transformation which interchanges quasiparticles with quasiholes. For a bilayer FQH system consisting
of two independent $1/m$ Laughlin FQH states, this includes the transformation which interchanges quasiparticles
from different layers. For $\mathbb{Z}_2$ quantum spin liquids, this includes electric-magnetic duality, 
which interchanges the $\mathbb{Z}_2$ gauge charges (spinons) with the $\mathbb{Z}_2$ fluxes.\cite{barkeshli2013genon} 

The action of a global symmetry group $G$ on the long wavelength effective TQFT is characterized first by a 
group homomorphism
\begin{align}
[\rho]: G \rightarrow \text{Aut}(\mathcal{C}) .
\end{align}
$[\rho_{\bf g}] \in \text{Aut}(\mathcal{C})$ describes how a given symmetry group element ${\bf g} \in G$ 
permutes the anyons of the system. $[\rho_{\bf g}]$ determines an action of $G$ such that all closed anyon 
diagrams in the UMTC are invariant; that is, all correlation functions of Wilson loop operators in the effective
TQFT description are invariant under the symmetry action. 

In Ref. \onlinecite{barkeshli2014SDG}, it was shown that despite the fact that $[\rho]$ appears to define an allowed symmetry
action for $G$ in the TQFT because all correlation functions will be invariant under $G$, the symmetry can have a certain severe anomaly. 
This anomaly is associated with the inability to localize the action of the symmetry to the location of the quasiparticles in a way that is consistent with associativity
of the group action. Consequently, we refer to this as a ``symmetry-localization'' anomaly. As we review
in the subsequent section, the map $[\rho]$ defines an element $[\coho{O}] \in \mathcal{H}^3_{[\rho]}(G, \mathcal{A})$. 
Here $\mathcal{A}$ is a finite Abelian group associated with the Abelian quasiparticles of $\mathcal{C}$, which form a 
group under fusion. $\mathcal{H}_{[\rho]}^3(G, \mathcal{A})$ is the third group cohomology of $G$ with coefficients in $\mathcal{A}$.
The subscript $[\rho]$ indicates that the cohomology depends on the action of $G$ on $\mathcal{A}$ through
$[\rho]$. 

As discussed in detail in Ref. \onlinecite{barkeshli2014SDG}, when $[\coho{O}]$ vanishes, then it is possible to consistently
define a notion of symmetry fractionalization. This specifies how quasiparticles carry fractional quantum numbers of the symmetry group
$G$; different possible symmetry fractionalization classes are related to each other by elements in 
$\mathcal{H}^2_{[\rho]}(G, \mathcal{A})$. However certain symmetry fractionalization classes may themselves be anomalous,
in the sense that they cannot exist in purely (2+1)D, but can exist at the surface of a (3+1)D SPT state.\cite{vishwanath2013, wang2013, 
MetlitskiPRB2013, Chen2014, Bonderson13d, chen2014b, YangPRL2015,hermele2016, song2016, wang2013b, metlitski2014, MetlitskiPRB2015, Fidkowski13,
SW1, ChoPRB2014, kapustin2014b} 
These may be referred to as anomalous symmetry fractionalization classes, or as ``SPT anomalies'' because of the
connection to the surface of (3+1)D SPTs. Using the language of the high energy field theory literature, these are examples of 't Hooft anomalies
in TQFTs. For space-time reflection symmetries which square to the identity, a general understanding of how to detect
such anomalies was presented in Ref. \onlinecite{barkeshli2016tr} by studying the theory on non-orientable space-time 
manifolds.\footnote{See also Ref. \onlinecite{wang2016,tachikawa2016a,tachikawa2016b} for related discussions 
that apply to the case of fermionic theories.} 
For unitary internal or lattice translation symmetries, a general understanding was developed in Ref. \onlinecite{barkeshli2014SDG,cheng2015} by solving 
consistency equations for the algebraic theory of symmetry defects.\footnote{For related references see also Ref. \onlinecite{Chen2014, WangPRX2016, cui2016}.}. 

For unitary internal (on-site) symmetries, known examples of the $\mathcal{H}^3$ symmetry localization anomaly 
occur for $G = \mathbb{Z}_2$ and are associated with discrete gauge theories with gauge group $\mathbb{D}_{20}$ or $\mathbb{D}_{16}$ 
(the dihedral groups with $20$ and $16$ elements, respectively).\cite{barkeshli2014SDG,fidkowski2015} Mathematically, 
the $\mathcal{H}^3$ obstructions in these examples have their roots in the theory of group extensions;\cite{brown1982} 
more recently, these obstructions appeared in the category theory literature, in the context of extending a 
fusion category by a group $G$.\cite{ENO2009} 

In this paper, we present and study in detail examples of the $\mathcal{H}^3$ anomaly for space-time parity 
odd symmetries, which include anti-unitary symmetries such as time-reversal, or unitary symmetries such as 
spatial reflections. Specifically, we consider cases where time-reversal symmetry ${\bf T}$ satisfies 
${\bf T}^2 = \mathds{1}$, or spatial reflection ${\bf R}$ satisfies ${\bf R}^2 = \mathds{1}$. We refer to these symmetry 
groups as $\mathbb{Z}_2^{\bf T}$ and $\mathbb{Z}_2^{\bf R}$, where the superscript denotes
the fact that the symmetry generator is anti-unitary or reverses the parity of space. 
These types of symmetries appear to be beyond what was considered in the relevant mathematical literature,
and the obstructions we find are not directly related to group extension obstructions of a gauge group, as in the previously known examples. 

The primary purpose of this work is to develop a deeper understanding of when and why the $\mathcal{H}^3$ 
symmetry localization anomaly occurs. As we review below, the $\mathcal{H}^3$ anomaly $[\coho{O}]$ can be explicitly computed 
from $[\rho]$ and the $F$ and $R$ symbols of $\mathcal{C}$, where the $F$ and $R$ symbols are certain consistent data that specify the
fusion and braiding properties of the anyons. However obtaining the $F$ and $R$ symbols of $\mathcal{C}$ 
is often tedious and computationally prohibitive. It is thus desirable to have a diagnostic for the presence
of the $\mathcal{H}^3$ anomaly from only the modular data of the theory (the modular $S$ matrix and topological spins). 
Here we provide such a set of constraints that, if violated, are sufficient to detect the existence of an $\mathcal{H}^3$
symmetry-localization anomaly. 

Recently, several infinite series of time-reversal invariant TQFTs were found by studying level-rank duality 
in Chern-Simons theories.\cite{aharony2016} The series of relevance to this paper are $\mr{USp}(2N)_N$ for $N$ even, 
and $\mr{SO}(N)_N$ for $N$ a multiple of $4$. We show that $\mr{USp}(4)_2$ and $\mr{SO}(4)_4$ 
are both part of an infinite family of theories with $\mathbb{Z}_2^{\bf T}$ symmetry-localization 
anomalies. We further show that in these theories, the anomaly can be resolved in several distinct ways:
(1) The presence of the $\mathcal{H}^3$ anomaly can be interpreted to mean that the symmetry of the theory was misidentified;
the true symmetry of the theory is the larger group, $\mathbb{Z}_4^{\bf T}$. We show in our examples that
$\mathbb{Z}_4^{\bf T}$ is free of the $\mathcal{H}^3$ symmetry localization anomaly. Therefore, these theories are 
time-reversal invariant, however ${\bf T}^2$ is a non-trivial unitary symmetry, while ${\bf T}^4 = \mathds{1}$. 
(2) We show that in the cases that we study, the anomaly can also be resolved by considering the TQFT
to be a theory of fermions (i.e. a spin TQFT), such that ${\bf T}^2 = (-1)^{N_f}$, where  $(-1)^{N_f}$
is the fermion parity of the system. 

While much of our discussion is phrased in terms of time-reversal symmetry, analogous results hold also
for reflection symmetry. To establish our results we will use whichever is convenient for the issues at hand,
noting that in the setup of Euclidean quantum field theory, which we make use of, time-reversal and 
spatial reflections appear on an equal footing. 

For the case where $G = \mathbb{Z}_2$ is a unitary internal (on-site) symmetry, Ref. \onlinecite{fidkowski2015} provided an 
example of an $\mathcal{H}^3$ anomaly that occurs for discrete gauge theory with gauge group $\mathbb{D}_{16}$. It was shown that 
it is possible to, in some sense, realize the theory at the surface of a (3+1)D SET with a global 
$G = \mathbb{Z}_2$ symmetry. In this case, the surface theory is no longer a (2+1)D theory, 
since some of the anyons that can exist on the surface correspond either to bulk quasiparticles
or to endpoints of strings in the (3+1)D bulk. For this reason, here we refer to this as a ``pseudo-realization''
of the original theory at the surface of the (3+1)D SET. In this construction, the $\mathcal{H}^3_{[\rho]}(G, \mathcal{A})$
could be related to the symmetry fractionalization of string excitations in the (3+1)D system~\cite{fidkowski2015, Cheng3D, ChenPRB2016, NingPRB2016}.

For the case where $G = \mathbb{Z}_2^{\bf T}$, symmetry fractionalization of string excitations in (3+1)D is not
well-understood, which makes the corresponding problem in this case especially intriguing. For the case of the
$\mathbb{Z}_2^{\bf T}$ symmetry localization anomalies in $\mr{USp}(4)_2$ and $\mr{SO}(4)_4$ CS theories, we demonstrate 
that the corresponding theories can be pseudo-realized at the surface of a (3+1)D SET, whose low energy
theory is a dynamical $\mathbb{Z}_2$ gauge theory and possesses a global $\mathbb{Z}_2^{\bf T}$
time-reversal symmetry. 

We further provide an infinite family of TQFTs with $\mathcal{H}^3_{[\rho]}(\mathbb{Z}_2^{\bf T}, \mathcal{A})$ symmetry
localization anomalies. These can all be obtained by starting with a theory with $\mathbb{Z}_4^{\bf T}$
symmetry, and gauging the unitary $\mathbb{Z}_2$ subgroup associated with ${\bf T}^2$.  {{ Importantly, when
performing the gauging of the unitary $\mathbb{Z}_2$ subgroup, we must add a Dijkgraaf-Witten term
for the $\mathbb{Z}_2$ gauge field, which is associated with the non-trivial element in 
$\mathcal{H}^3(\mathbb{Z}_2, \mathrm{U}(1)) = \mathbb{Z}_2$. Physically, this corresponds to stacking a $\mathbb{Z}_2$
SPT with the theory with the $\mathbb{Z}_4^{{\bf T}}$ symmetry.}} This process can be repeated
indefinitely, so that we can generate an infinite series of TQFTs with $\mathcal{H}^3_{[\rho]}(\mathbb{Z}_2^{\bf T}, \mathcal{A})$ 
anomalies by starting with a single ``root'' theory with $\mathbb{Z}_4^{\bf T}$ symmetry. We provide an infinite set of 
such ``root''  theories. For the USp$(4)_2$ and SO$(4)_4$ examples, the root phases are SU$(5)_1$ and SU$(3)_1 \times$ SU$(3)_1$ 
CS theory, respectively. We show that $\text{U}(1) \times \text{U}(1)$ CS theory with $K$-matrix $K = \left( \begin{matrix}
 m & n \\ n & -m \end{matrix} \right)$ provides another infinite family of root theories as well. 

This paper is organized as follows. In Sec. \ref{H3review} we provide a brief review of UMTCs and the action of
global symmetry, the general definition of the $\mathcal{H}^3_{[\rho]}(G,\mathcal{A})$ anomaly, and symmetry fractionalization. 
In Sec. \ref{UspSec} we introduce the example of USp$(4)_2$ CS theory as a theory with a $\mathcal{H}^3_{[\rho]}(\mathbb{Z}_2^{\bf T},\mathcal{A})$
anomaly. In Sec. \ref{constraints} we provide a set of sufficient conditions, defined in terms of the modular data of the theory and
the way ${\bf T}$ permutes the anyons, for diagnosing $\mathcal{H}^3_{[\rho]}(\mathbb{Z}_2^{\bf T},\mathcal{A})$ anomalies. 
In Sec. \ref{resolutions} we discuss several resolutions of the $\mathcal{H}^3$ anomaly, which include enlarging the symmetry from
$\mathbb{Z}_2^{\bf T}$ to $\mathbb{Z}_4^{\bf T}$, viewing the theory as a theory of fermions, and finally how to 
pseudo-realize theories with such anomalies at the surface of (3+1)D SET phases. In Sec. \ref{examples} we provide an 
infinite family of examples that generalize the USp$(4)_2$ example. We conclude with a discussion of some open issues 
in Sec. \ref{disc}. 

\section{Review of the $\mathcal{H}^3_{[\rho]}(G,\mathcal{A})$ anomaly}
\label{H3review}

In this section we briefly review the discussion of Ref. \onlinecite{barkeshli2014SDG} regarding the 
$\mathcal{H}^3_{[\rho]}(G,\mathcal{A})$ symmetry localization anomaly. 

\subsection{Review of UMTC notation}

Here we briefly review the notation that we use to describe UMTCs. For a more comprehensive review
of the notation that we use, see e.g. Ref. \onlinecite{barkeshli2014SDG}. The topologically 
non-trivial quasiparticles of a (2+1)D topologically ordered state are equivalently referred to
as anyons, topological charges, and quasiparticles. In the category theory terminology, they correspond
to isomorphism classes of simple objects of the UMTC. 

A UMTC $\mathcal{C}$ contains splitting spaces $V_{c}^{ab}$, and their dual fusion spaces, $V_{ab}^c$,
where $a,b,c \in \mathcal{C}$ are the anyons. These spaces have dimension 
$\text{dim } V_{c}^{ab} = \text{dim } V_{ab}^c = N_{ab}^c$, where $N_{ab}^c$ are referred
to as the fusion rules. They are depicted graphically as: 
\begin{equation}
\left( d_{c} / d_{a}d_{b} \right) ^{1/4}
\pspicture[shift=-0.6](-0.1,-0.2)(1.5,-1.2)
  \small
  \psset{linewidth=0.9pt,linecolor=black,arrowscale=1.5,arrowinset=0.15}
  \psline{-<}(0.7,0)(0.7,-0.35)
  \psline(0.7,0)(0.7,-0.55)
  \psline(0.7,-0.55) (0.25,-1)
  \psline{-<}(0.7,-0.55)(0.35,-0.9)
  \psline(0.7,-0.55) (1.15,-1)	
  \psline{-<}(0.7,-0.55)(1.05,-0.9)
  \rput[tl]{0}(0.4,0){$c$}
  \rput[br]{0}(1.4,-0.95){$b$}
  \rput[bl]{0}(0,-0.95){$a$}
 \scriptsize
  \rput[bl]{0}(0.85,-0.5){$\mu$}
  \endpspicture
=\left\langle a,b;c,\mu \right| \in
V_{ab}^{c} ,
\label{eq:bra}
\end{equation}
\begin{equation}
\left( d_{c} / d_{a}d_{b}\right) ^{1/4}
\pspicture[shift=-0.65](-0.1,-0.2)(1.5,1.2)
  \small
  \psset{linewidth=0.9pt,linecolor=black,arrowscale=1.5,arrowinset=0.15}
  \psline{->}(0.7,0)(0.7,0.45)
  \psline(0.7,0)(0.7,0.55)
  \psline(0.7,0.55) (0.25,1)
  \psline{->}(0.7,0.55)(0.3,0.95)
  \psline(0.7,0.55) (1.15,1)	
  \psline{->}(0.7,0.55)(1.1,0.95)
  \rput[bl]{0}(0.4,0){$c$}
  \rput[br]{0}(1.4,0.8){$b$}
  \rput[bl]{0}(0,0.8){$a$}
 \scriptsize
  \rput[bl]{0}(0.85,0.35){$\mu$}
  \endpspicture
=\left| a,b;c,\mu \right\rangle \in
V_{c}^{ab},
\label{eq:ket}
\end{equation}
where $\mu=1,\ldots ,N_{ab}^{c}$, $d_a$ is the quantum dimension of $a$, 
and the factors $\left(\frac{d_c}{d_a d_b}\right)^{1/4}$ are a normalization convention for the diagrams. 

We denote $\bar{a}$ as the topological charge conjugate of $a$, for which
$N_{a \bar{a}}^1 = 1$, i.e.
\begin{align}
a \times \bar{a} = 1 +\cdots
\end{align}
Here $1$ refers to the identity particle, i.e. the vacuum topological sector, which physically describes all 
local, topologically trivial excitations. 

The $F$-symbols are defined as the following basis transformation between the splitting
spaces of $4$ anyons:
\begin{equation}
  \pspicture[shift=-1.0](0,-0.45)(1.8,1.8)
  \small
  \psset{linewidth=0.9pt,linecolor=black,arrowscale=1.5,arrowinset=0.15}
  \psline(0.2,1.5)(1,0.5)
  \psline(1,0.5)(1,0)
  \psline(1.8,1.5) (1,0.5)
  \psline(0.6,1) (1,1.5)
   \psline{->}(0.6,1)(0.3,1.375)
   \psline{->}(0.6,1)(0.9,1.375)
   \psline{->}(1,0.5)(1.7,1.375)
   \psline{->}(1,0.5)(0.7,0.875)
   \psline{->}(1,0)(1,0.375)
   \rput[bl]{0}(0.05,1.6){$a$}
   \rput[bl]{0}(0.95,1.6){$b$}
   \rput[bl]{0}(1.75,1.6){${c}$}
   \rput[bl]{0}(0.5,0.5){$e$}
   \rput[bl]{0}(0.9,-0.3){$d$}
 \scriptsize
   \rput[bl]{0}(0.3,0.8){$\alpha$}
   \rput[bl]{0}(0.7,0.25){$\beta$}
  \endpspicture
= \sum_{f,\mu,\nu} \left[F_d^{abc}\right]_{(e,\alpha,\beta)(f,\mu,\nu)}
 \pspicture[shift=-1.0](0,-0.45)(1.8,1.8)
  \small
  \psset{linewidth=0.9pt,linecolor=black,arrowscale=1.5,arrowinset=0.15}
  \psline(0.2,1.5)(1,0.5)
  \psline(1,0.5)(1,0)
  \psline(1.8,1.5) (1,0.5)
  \psline(1.4,1) (1,1.5)
   \psline{->}(0.6,1)(0.3,1.375)
   \psline{->}(1.4,1)(1.1,1.375)
   \psline{->}(1,0.5)(1.7,1.375)
   \psline{->}(1,0.5)(1.3,0.875)
   \psline{->}(1,0)(1,0.375)
   \rput[bl]{0}(0.05,1.6){$a$}
   \rput[bl]{0}(0.95,1.6){$b$}
   \rput[bl]{0}(1.75,1.6){${c}$}
   \rput[bl]{0}(1.25,0.45){$f$}
   \rput[bl]{0}(0.9,-0.3){$d$}
 \scriptsize
   \rput[bl]{0}(1.5,0.8){$\mu$}
   \rput[bl]{0}(0.7,0.25){$\nu$}
  \endpspicture
.
\end{equation}
To describe topological phases, these are required to be unitary transformations, i.e.
\begin{eqnarray}
\left[ \left( F_{d}^{abc}\right) ^{-1}\right] _{\left( f,\mu
,\nu \right) \left( e,\alpha ,\beta \right) }
&=& \left[ \left( F_{d}^{abc}\right) ^{\dagger }\right]
_{\left( f,\mu ,\nu \right) \left( e,\alpha ,\beta \right) }
\notag \\
&=& \left[ F_{d}^{abc}\right] _{\left( e,\alpha ,\beta \right) \left( f,\mu
,\nu \right) }^{\ast }
.
\end{eqnarray}

The $R$-symbols define the braiding properties of the anyons, and are defined via the the following
diagram:
\begin{equation}
\pspicture[shift=-0.65](-0.1,-0.2)(1.5,1.2)
  \small
  \psset{linewidth=0.9pt,linecolor=black,arrowscale=1.5,arrowinset=0.15}
  \psline{->}(0.7,0)(0.7,0.43)
  \psline(0.7,0)(0.7,0.5)
 \psarc(0.8,0.6732051){0.2}{120}{240}
 \psarc(0.6,0.6732051){0.2}{-60}{35}
  \psline (0.6134,0.896410)(0.267,1.09641)
  \psline{->}(0.6134,0.896410)(0.35359,1.04641)
  \psline(0.7,0.846410) (1.1330,1.096410)	
  \psline{->}(0.7,0.846410)(1.04641,1.04641)
  \rput[bl]{0}(0.4,0){$c$}
  \rput[br]{0}(1.35,0.85){$b$}
  \rput[bl]{0}(0.05,0.85){$a$}
 \scriptsize
  \rput[bl]{0}(0.82,0.35){$\mu$}
  \endpspicture
=\sum\limits_{\nu }\left[ R_{c}^{ab}\right] _{\mu \nu}
\pspicture[shift=-0.65](-0.1,-0.2)(1.5,1.2)
  \small
  \psset{linewidth=0.9pt,linecolor=black,arrowscale=1.5,arrowinset=0.15}
  \psline{->}(0.7,0)(0.7,0.45)
  \psline(0.7,0)(0.7,0.55)
  \psline(0.7,0.55) (0.25,1)
  \psline{->}(0.7,0.55)(0.3,0.95)
  \psline(0.7,0.55) (1.15,1)	
  \psline{->}(0.7,0.55)(1.1,0.95)
  \rput[bl]{0}(0.4,0){$c$}
  \rput[br]{0}(1.4,0.8){$b$}
  \rput[bl]{0}(0,0.8){$a$}
 \scriptsize
  \rput[bl]{0}(0.82,0.37){$\nu$}
  \endpspicture
.
\end{equation}

The topological twist $\theta_a = e^{2\pi i h_a}$, with $h_a$ the topological spin, is defined
via the diagram:
\begin{equation}
\theta _{a}=\theta _{\bar{a}}
=\sum\limits_{c,\mu } \frac{d_{c}}{d_{a}}\left[ R_{c}^{aa}\right] _{\mu \mu }
= \frac{1}{d_{a}}
\pspicture[shift=-0.5](-1.3,-0.6)(1.3,0.6)
\small
  \psset{linewidth=0.9pt,linecolor=black,arrowscale=1.5,arrowinset=0.15}
  \psarc[linewidth=0.9pt,linecolor=black] (0.7071,0.0){0.5}{-135}{135}
  \psarc[linewidth=0.9pt,linecolor=black] (-0.7071,0.0){0.5}{45}{315}
  \psline(-0.3536,0.3536)(0.3536,-0.3536)
  \psline[border=2.3pt](-0.3536,-0.3536)(0.3536,0.3536)
  \psline[border=2.3pt]{->}(-0.3536,-0.3536)(0.0,0.0)
  \rput[bl]{0}(-0.2,-0.5){$a$}
  \endpspicture
,
\end{equation}

Finally, the modular, or topological, $S$-matrix, is defined as
\begin{equation}
S_{ab} =\mathcal{D}^{-1}\sum%
\limits_{c}N_{\bar{a} b}^{c}\frac{\theta _{c}}{\theta _{a}\theta _{b}}d_{c}
=\frac{1}{\mathcal{D}}
\pspicture[shift=-0.4](0.0,0.2)(2.6,1.3)
\small
  \psarc[linewidth=0.9pt,linecolor=black,arrows=<-,arrowscale=1.5,arrowinset=0.15] (1.6,0.7){0.5}{167}{373}
  \psarc[linewidth=0.9pt,linecolor=black,border=3pt,arrows=<-,arrowscale=1.5,arrowinset=0.15] (0.9,0.7){0.5}{167}{373}
  \psarc[linewidth=0.9pt,linecolor=black] (0.9,0.7){0.5}{0}{180}
  \psarc[linewidth=0.9pt,linecolor=black,border=3pt] (1.6,0.7){0.5}{45}{150}
  \psarc[linewidth=0.9pt,linecolor=black] (1.6,0.7){0.5}{0}{50}
  \psarc[linewidth=0.9pt,linecolor=black] (1.6,0.7){0.5}{145}{180}
  \rput[bl]{0}(0.1,0.45){$a$}
  \rput[bl]{0}(0.8,0.45){$b$}
  \endpspicture
,
\label{eqn:mtcs}
\end{equation}
where $\mathcal{D} = \sqrt{\sum_a d_a^2}$.

\subsection{Topological symmetry and braided auto-equivalence}

An important property of a UMTC $\mathcal{C}$ is the group of ``topological symmetries,'' 
which are related to ``braided auto-equivalences'' in the mathematical
literature. They are associated with the symmetries of the emergent TQFT described by $\mathcal{C}$, irrespective
of any microscopic global symmetries of a quantum system in which the TQFT emerges as the long wavelength
description. 

The topological symmetries consist of the invertible maps
\begin{align}
\varphi: \mathcal{C} \rightarrow \mathcal{C} .
\end{align}
The different $\varphi$, modulo equivalences known as natural isomorphisms, form a group, which we denote
as Aut$(\mathcal{C})$.\cite{barkeshli2014SDG}

The symmetry maps can be classified according to a $\mathbb{Z}_2 \times \mathbb{Z}_2$ grading,
defined by 
\begin{align}
q(\varphi) = \left\{
\begin{array} {ll}
0 & \text{if $\varphi$ is not time-reversing} \\
1 & \text{if $\varphi$ is time-reversing} \\
\end{array} \right.
\end{align}
\begin{align}
p(\varphi) = \left\{
\begin{array} {ll}
0 & \text{if $\varphi$ is spatial parity even} \\
1 & \text{if $\varphi$ is spatial parity odd} \\
\end{array} \right.
\end{align}
Here time-reversing transformations are anti-unitary, while spatial parity odd transformations
involve an odd number of reflections in space, thus changing the orientation of space. Thus the
topological symmetry group can be decomposed as
\begin{align}
\text{Aut}(\mathcal{C}) = \bigsqcup_{q,p=0,1} \text{Aut}_{q,p}(\mathcal{C}) .
\end{align}
Aut$_{0,0}(\mathcal{C})$ is therefore the subgroup corresponding to topological symmetries that are 
unitary and space-time parity even (this is referred to in the mathematical literature as the group of
``braided auto-equivalences''). The generalization involving reflection and time-reversal symmetries
appears to be beyond what has been considered in the mathematics literature to date. 

It is also convenient to define
\begin{align}
\sigma(\varphi) = \left\{
\begin{array} {ll}
1 & \text{if $\varphi$ is space-time parity even} \\
* & \text{if $\varphi$ is space-time parity odd} \\
\end{array} \right.
\end{align}
A map $\varphi$ is space-time parity odd if $(q(\varphi) + p(\varphi)) \text{ mod } 2 = 1$, and otherwise
it is space-time parity even. 

The maps $\varphi$ may permute the topological charges:
\begin{align}
\varphi(a) = a' \in \mathcal{C}, 
\end{align}
subject to the constraint that 
\begin{align}
N_{a'b'}^{c'} &= N_{ab}^c
\nonumber \\
S_{a'b'} &= S_{ab}^{\sigma(\varphi)},
\nonumber \\
\theta_{a'} &= \theta_a^{\sigma(\varphi)},
\end{align}
The maps $\varphi$ have a corresponding action on the $F$- and $R-$ symbols of the theory,
as well as on the fusion and splitting spaces, which we will discuss in the subsequent section. 

\subsection{Global symmetry}

Let us now suppose that we are interested in a system with a global symmetry group $G$. For example, we may be interested
in a given microscopic Hamiltonian that has a global symmetry group $G$, whose ground state preserves $G$, and whose 
anyonic excitations are algebraically described by $\mathcal{C}$. The global symmetry acts on the topological quasiparticles 
and the topological state space through the action of a group homomorphism
\begin{align}
[\rho] : G \rightarrow \text{Aut}(\mathcal{C}) . 
\end{align}
We use the notation $[\rho_{\bf g}] \in \text{Aut}(\mathcal{C})$ for a specific element ${\bf g} \in G$. The square
brackets indicate the equivalence class of symmetry maps related by natural isomorphisms, which we define below. $\rho_{\bf g}$ is thus a
representative symmetry map of the equivalence class $[\rho_{\bf g}]$. We use the notation
\begin{align}
\,^{\bf g}a \equiv \rho_{\bf g}(a). 
\end{align}
We associate gradings $q({\bf g})$ and $p({\bf g})$ by defining
\begin{align}
q({\bf g}) &\equiv q(\rho_{\bf g}) 
\nonumber \\
p({\bf g}) &\equiv p(\rho_{\bf g}) 
\nonumber \\
\sigma({\bf g}) &\equiv \sigma( \rho_{\bf g})
\end{align}

$\rho_{\bf g}$ has an action on the fusion/splitting spaces:
\begin{align}
\rho_{\bf g} : V_{ab}^c \rightarrow V_{\,^{\bf g}a \,^{\bf g}b}^{\,^{\bf g}c} .   
\end{align}
This map is unitary if $q({\bf g}) = 0$ and anti-unitary if $q({\bf g}) = 1$. We write this as
\begin{align}
\rho_{\bf g} |a,b;c, \mu\rangle = \sum_{\nu} [U_{\bf g}(\,^{\bf g}a , \,^{\bf g}b ; \,^{\bf g}c )]_{\mu\nu} K^{q({\bf g})} |a,b;c,\nu\rangle,
\end{align}
where $U_{\bf g}(\,^{\bf g}a , \,^{\bf g}b ; \,^{\bf g}c ) $ is a $N_{ab}^c \times N_{ab}^c$ matrix, and 
$K$ denotes complex conjugation.\footnote{We note that for spatial reflection symmetries,
we also must consider the corresponding maps $\rho_{\bf g} : V_{ab}^c \rightarrow V_{\,^{\bf g}b \,^{\bf g}a}^{\,^{\bf g}c}$, 
depending on whether we consider the anyons to be aligned perpendicular to the reflection axis or along with it. }

Under the map $\rho_{\bf g}$, the $F$ and $R$ symbols transform as well:
\begin{widetext}
\begin{align}
\rho_{\bf g}[ F^{abc}_{def}] &= U_{\bf g}(\,^{\bf g}a, \,^{\bf g}b; \,^{\bf g}e) U_{\bf g}(\,^{\bf g}e, \,^{\bf g}c; \,^{\bf g}d) F^{\,^{\bf g}a \,^{\bf g}b \,^{\bf g}c }_{\,^{\bf g}d \,^{\bf g}e \,^{\bf g}f} 
U^{-1}_{\bf g}(\,^{\bf g}b, \,^{\bf g}c; \,^{\bf g}f) U^{-1}_{\bf g}(\,^{\bf g}a, \,^{\bf g}f; \,^{\bf g}d) = K^{\sigma({\bf g})} F^{abc}_{def} K^{\sigma({\bf g})}
\nonumber \\
\rho_{\bf g} [R^{ab}_c] &= U_{\bf g}(\,^{\bf g}a, \,^{\bf g}b; \,^{\bf g}c)  R^{\,^{\bf g}a \,^{\bf g}b}_{\,^{\bf g}c} U_{\bf g}(\,^{\bf g}a, \,^{\bf g}b; \,^{\bf g}c)^{-1} = K^{\sigma({\bf g})} R^{ab}_c K^{\sigma({\bf g})},
\end{align}
\end{widetext}
where we have suppressed the additional indices that appear when $N_{ab}^c > 1$. 

Importantly, we have
\begin{align}
\kappa_{{\bf g}, {\bf h}} \circ \rho_{\bf g} \circ \rho_{\bf h} = \rho_{\bf g h} ,
\end{align}
where the action of $\kappa_{ {\bf g}, {\bf h}}$ on the fusion / splitting spaces is defined as
\begin{align}
\kappa_{ {\bf g}, {\bf h}} ( |a, b;c,\mu \rangle) = \sum_\nu [\kappa_{ {\bf g}, {\bf h}} ( a, b;c )]_{\mu\nu} | a, b;c,\nu \rangle.
\end{align}
The above definitions imply that
\begin{widetext}
\begin{align}
\label{kappaU}
\kappa_{ {\bf g}, {\bf h}} ( a, b;c ) = U_{\bf g}(a,b;c)^{-1} K^{q({\bf g})} U_{\bf h}( \,^{\bar{\bf g}}a, \,^{\bar{\bf g}}b; \,^{\bar{\bf g}}c  )^{-1} K^{q({\bf q})} U_{\bf gh}(a,b;c ),
\end{align}
\end{widetext}
where $\bar{\bf g} \equiv {\bf g}^{-1}$. $\kappa_{ {\bf g}, {\bf h}}$ is a natural isomorphism, which means that by definition,
\begin{align}
[\kappa_{ {\bf g}, {\bf h}} (a,b;c)]_{\mu \nu} = \delta_{\mu \nu} \frac{\beta_a({\bf g}, {\bf h}) \beta_b({\bf g}, {\bf h})}{\beta_c({\bf g}, {\bf h}) },
\end{align}
where $\beta_a({\bf g}, {\bf h})$ are $U(1)$ phases. 

\subsection{$\mathcal{H}_{[\rho]}^3(G, \mathcal{A})$ obstruction}

As discussed in detail in Ref. \onlinecite{barkeshli2014SDG}, the choice of $[\rho_{\bf g}]$ defines an element 
$[\coho{O}] \in \mathcal{H}^3_{[\rho]}(G, \mathcal{A})$. To see this, we first define
\begin{align}
\Omega_a({\bf g}, {\bf h}, {\bf k}) = \frac{K^{\sigma({\bf g})} \beta_{\rho_{\bf g}^{-1}(a)}({\bf h}, {\bf k}) K^{\sigma({\bf g})} \beta_a({\bf g}, {\bf h k})}{\beta_a({\bf g h}, {\bf k}) \beta_a({\bf g}, {\bf h})} 
\end{align}
It can be shown that
\begin{align}
\Omega_a({\bf g}, {\bf h}, {\bf k}) \Omega_b({\bf g}, {\bf h}, {\bf k}) = \Omega_c({\bf g}, {\bf h}, {\bf k}) ,
\end{align}
if $N_{ab}^c \neq 0$. This then implies that\cite{barkeshli2014SDG}
\begin{align}
\Omega_a({\bf g}, {\bf h}, {\bf k}) = M_{a\cohosub{O}({\bf g}, {\bf h}, {\bf k})}^* ,
\end{align}
for some $\coho{O}({\bf g}, {\bf h}, {\bf k}) \in \mathcal{A}$. Here $\mathcal{A} \subset \mathcal{C}$ is the subset of topological charges
in $\mathcal{C}$ that are Abelian. These form a finite group, which we also denote $\mathcal{A}$, under fusion. 
Given an Abelian anyon $b \in \mathcal{A}$, $M_{ab}$ is the braiding phase obtained by encircling $b$ around $a$, as defined by the
following anyon diagram:
\begin{equation}
\label{Mabdef}
  \pspicture[shift=-0.6](0.0,-0.05)(1.1,1.45)
  \small
  \psarc[linewidth=0.9pt,linecolor=black,border=0pt] (0.8,0.7){0.4}{120}{225}
  \psarc[linewidth=0.9pt,linecolor=black,arrows=<-,arrowscale=1.4,
    arrowinset=0.15] (0.8,0.7){0.4}{165}{225}
  \psarc[linewidth=0.9pt,linecolor=black,border=0pt] (0.4,0.7){0.4}{-60}{45}
  \psarc[linewidth=0.9pt,linecolor=black,arrows=->,arrowscale=1.4,
    arrowinset=0.15] (0.4,0.7){0.4}{-60}{15}
  \psarc[linewidth=0.9pt,linecolor=black,border=0pt]
(0.8,1.39282){0.4}{180}{225}
  \psarc[linewidth=0.9pt,linecolor=black,border=0pt]
(0.4,1.39282){0.4}{-60}{0}
  \psarc[linewidth=0.9pt,linecolor=black,border=0pt]
(0.8,0.00718){0.4}{120}{180}
  \psarc[linewidth=0.9pt,linecolor=black,border=0pt]
(0.4,0.00718){0.4}{0}{45}
  \rput[bl]{0}(0.1,1.2){$a$}
  \rput[br]{0}(1.06,1.2){$b$}
  \endpspicture
= M_{ab}
\pspicture[shift=-0.6](-0.2,-0.45)(1.0,1.1)
  \small
  \psset{linewidth=0.9pt,linecolor=black,arrowscale=1.5,arrowinset=0.15}
  \psline(0.3,-0.4)(0.3,1)
  \psline{->}(0.3,-0.4)(0.3,0.50)
  \psline(0.7,-0.4)(0.7,1)
  \psline{->}(0.7,-0.4)(0.7,0.50)
  \rput[br]{0}(0.96,0.8){$b$}
  \rput[bl]{0}(0,0.8){$a$}
  \endpspicture
.
\end{equation}
In terms of the modular $S$-matrix, $M_{ab} = \frac{S_{ab}^* S_{00}}{S_{0a} S_{0b}}$.

One can show that $\coho{O}$ is a 3-cocycle, and that there is a freedom in the choice of $\beta_a$ 
which relate two different $\coho{O}$'s by a 3-coboundary. Therefore $[\rho_{\bf g}]$ defines an element in 
$[\coho{O}] \in \mathcal{H}^3_{[\rho]}(G, \mathcal{A})$. 

In Ref. \onlinecite{barkeshli2014SDG}, it was shown that $[\coho{O}] \in \mathcal{H}^3_{[\rho]}(G, \mathcal{A})$ is an 
obstruction to symmetry localization. That is, it is an obstruction to consistently defining a notion of symmetry action on individual 
topological charges. More specifically, let us consider a state $|\Psi_{a_1, \cdots, a_n} \rangle$ in the full Hilbert space of the system, 
which consists of $n$ anyons, $a_1, \cdots a_n$, at well-separated locations, which collectively fuse to the identity topological sector. 
Since the ground state is $G$-symmetric, we expect that the symmetry action $R_{\bf g}$ on this state decomposes as follows:
\begin{align}
\label{symloc}
R_{\bf g}  |\Psi_{a_1, \cdots, a_n} \rangle \approx \prod_{j = 1}^n U^{(j)}_{\bf g} U_{\bf g} (\,^{\bf g}a_1, \cdots, \,^{\bf g}a_n ; 0) |\Psi_{\,^{\bf g}a_1, \cdots, \,^{\bf g}a_n} \rangle .
\end{align}
Here, $U^{(j)}_{\bf g}$ are unitary matrices that have support in a region (of length scale set by the correlation length)
localized to the anyon $a_j$. The map $U_{\bf g} (\,^{\bf g}a_1, \cdots, \,^{\bf g}a_n ; 0)$ is the generalization of
$U_{\bf g} (\,^{\bf g}a,\,^{\bf g} b; \,^{\bf g} c)$, defined above, to the case with $n$ anyons fusing to vacuum. In contrast to the
local unitaries $U^{(j)}_{\bf g}$, $U_{\bf g} (\,^{\bf g}a_1, \cdots, \,^{\bf g}a_n ; 0)$ only depends on the global topological sector of the system (i.e. on the 
precise fusion tree that defines the topological state). $R_{\bf g}$ is the representation of ${\bf g}$ acting
on the full Hilbert space of the theory. The $\approx$ means that the equation is true up to corrections that are exponentially small
in the distance between the anyons and the correlation length of the system. 
$[\coho{O}] \in \mathcal{H}^3_{[\rho]}(G, \mathcal{A})$ is an obstruction to
Eq. (\ref{symloc}) being consistent when considering the associativity of three group elements.\cite{barkeshli2014SDG}.  

When $\rho_{\bf g}$ does not permute any anyons, i.e. $\rho_{\bf g}(a) = a$ for all $a$, we expect
that $\rho_{\bf g}$ must be a natural isomorphism. This has so far been proven rigorously 
for the case where $\mathcal{C}$ is an Abelian theory. One can show that this implies that the associated 
$\mathcal{H}^3$ obstruction is always vanishing. With this assumption, non-vanishing $\mathcal{H}^3$ 
obstructions therefore require $\rho_{\bf g}$ to have a non-trivial permutation action on the anyons. 

\subsection{Symmetry fractionalization}

When the $\mathcal{H}^3_{[\rho]}(G, \mathcal{A})$ symmetry-localization obstruction vanishes, then one can define
a consistent notion of symmetry fractionalization. Symmetry fractionalization determines how the anyons in the system
carry fractional symmetry quantum numbers. In general, the distinct allowed patterns of symmetry fractionalization are
in one-to-one correspondence with elements in $\mathcal{H}^2_{[\rho]}(G, \mathcal{A})$.\cite{barkeshli2014SDG} 

For the case of time-reversal symmetry, an important set of data that characterizes time-reversal symmetry
fractionalization is as follows. When $a = \,^{\bf T}a$, one can define a quantity $\eta_a^{\bf T} = \pm 1$.
This determines whether locally the action of ${\bf T}^2$ on $a$ is equal to $\pm 1$.\cite{LevinPRB2012,barkeshli2014SDG} This, in turn,
determines whether $a$ carries a ``local Kramers degeneracy.'' If $\eta_a^{\bf T} = -1$, then $a$ carries with it an 
internal local multi-dimensional Hilbert space whose degeneracy (the Kramers degeneracy) 
is protected by time-reversal symmetry. The quantities $\eta_a^{\bf T}$ must satisfy a number of highly non-trivial
consistency relations. For example, one can show:\cite{barkeshli2014SDG}
\begin{align}
\label{kappaEtaEq}
[\kappa_{{\bf T}}(a,b;c)]_{\mu \nu} = \delta_{\mu \nu} \frac{\eta_a^{\bf T} \eta_b^{\bf T}}{\eta_c^{\bf T}} . 
\end{align}
Moreover, if $N_{ab}^c$ is odd and $\,^{\bf T}a = a$, $\,^{\bf T}b = b$, and 
$\,^{\bf T}c = c$, then 
\begin{align}
\label{etaFusionEq}
\eta_a^{\bf T} \eta_b^{\bf T} = \eta_c^{\bf T}. 
\end{align}
Similarly, if $N_{a \,^{\bf T} a}^b$ is odd and $\,^{\bf T}b = b$ one can prove
\begin{align}
\label{etaTwistEq}
\eta_b^{\bf T} = \theta_b .
\end{align}

The case of reflection symmetry is analogous to that of time-reversal symmetry. Here, when $a = \,^{\bf R} \bar{a}$, 
then we can define a symmetry fractionalization quantum number $\eta_{a}^{\bf R} = \pm 1$.
$\eta_a^{\bf R}$ can be understood as follows. We place $a$ and $\bar{a}$ away from each other, such that the action
of reflection ${\bf R}$ interchanges their positions. If $a = \,^{\bf R} \bar{a}$, then the system is reflection invariant, and
$\eta_a^{\bf R} = \pm 1$ corresponds to the eigenvalue of the action of reflection on this state. Alternatively, 
we can consider taking the spatial manifold to be a cylinder, with topological
charge $a$ and $\bar{a}$ on the two ends of the cylinder, such that the action of reflection interchanges their position. 
If $a = \,^{\bf R}\bar{a}$, then we can view this system as a (1+1)D reflection symmetry SPT system, which has a $\mathbb{Z}_2$ 
classification. $\eta_a^{\bf R}$ can be related to whether this (1+1)D reflection SPT is trivial or non-trivial (when compared to
the case where $a$ is the identity particle). See Ref. \onlinecite{barkeshli2016tr} for more details. 

\section{Example: $\mr{USp}(4)_2$ Chern-Simons Theory}
\label{UspSec}

An explicit example of a theory with a $\mathcal{H}^3(\mathbb{Z}_2^{\bf T}, \mathcal{A})$ anomaly
is Chern-Simons theory with gauge group $\mr{USp}(4)_2$. Here $\mr{USp}(2n)$ is the symplectic group,
where $\mr{USp}(2) = \mr{SU}(2)$.\footnote{Note that $\mr{USp}(2n) = \mr{Sp}(n)$, where $\mr{Sp}(1) = \mr{SU}(2)$.} 
The anyon content of $\mr{USp}(4)_2$ CS theory coincides with the 
integrable highest weight representations of the affine lie aglebra $\mathfrak{so}(5)_2$.\cite{witten1989,difrancesco} It consists of $6$ particles, 
which we can label $1, \epsilon, \phi_1, \phi_2, \psi_+, \psi_-$. 
 The fusion rules are given by
\begin{equation}
	\begin{gathered}
	\epsilon\times\epsilon=1, \;\; \epsilon\times \phi_i=\phi_i,\;\; \epsilon\times\psi_+ = \psi_-\\
	\phi_i\times\phi_i=1+\epsilon+\phi_{\min(2i, 5-2i)}, \\
 \phi_1\times\phi_2=\phi_1+\phi_2\\ 
	\psi_+\times\psi_+=1+\phi_1+\phi_2.
	\end{gathered}
	\label{}
\end{equation}
Here $i=1,2$. We also list the quantum dimensions and topological twists in Table \ref{tab:so52}, 
from which one can construct the modular $S$ matrix:
\begin{align}
S = \frac{1}{\sqrt{20}} \left( 
\begin{matrix}
1 & 1 & 2 & \sqrt{5} & \sqrt{5} & 2 \\
1 & 1 & 2 & -\sqrt{5} &-\sqrt{5} & 2  \\
2 & 2 & \sqrt{5}-1 & 0 & 0  & -\sqrt{5}-1  \\
\sqrt{5} & -\sqrt{5} & 0 & \sqrt{5} & - \sqrt{5} & 0 \\
\sqrt{5} & -\sqrt{5}  & 0 & -\sqrt{5} & \sqrt{5} & 0 \\
2 & 2  & -\sqrt{5} - 1 & 0 & 0 & \sqrt{5}-1\\
\end{matrix} \right),
\end{align}
where we have presented $S$ in the basis $(1,\epsilon, \phi_1, \psi_+, \psi_-, \phi_2)$. 

The $F$ and $R$ symbols of this theory are tabulated in Ref. \onlinecite{ardonne2016}. 
\begin{table}
	\centering
	\begin{tabular}{c|c|c|c|c|c|c}
		\hline
		Anyon label, $a$ & $1$ & $\varepsilon$ & $\phi_1$ & $\phi_2$ & $\psi_+$ & $\psi_-$\\
		\hline
		Quantum dimension, $d_a$ & 1 & 1 & 2 & 2 & $\sqrt{5}$ & $\sqrt{5}$\\
		\hline
		Topological twist, $\theta_a$ & 1 & 1 & $e^{\frac{4\pi i}{5}}$ & $e^{-\frac{4\pi i}{5}}$ & $i$ & $-i$\\
		\hline
		Time-reversal action, $\,^{\bf T}a$ & 1 & $\varepsilon$ & $\phi_2$ & $\phi_1$ & $\psi_-$ & $\psi_+$\\
		\hline
	\end{tabular}
	\caption{Anyon content of USp(4)$_2$.}
	\label{tab:so52}
\end{table}
We see that there is only one possible action under time-reversal, summarized in Table \ref{tab:so52}.

In this case, $\mathcal{A} = \mathbb{Z}_2$, and $\H^3(\mathbb{Z}_2^{\bf T}, \mathbb{Z}_2) = \mathbb{Z}_2$. 
By direct computation, following the procedure outlined in the previous section, we find that the choice of $[\rho_{\bf T}]$ described above leads
to a non-trivial obstruction $[\coho{O}] \in \mathcal{H}^3(\mathbb{Z}_2^{\bf T}, \mathbb{Z}_2) $. In fact, one finds that the representative $3$-cocycle is given by $\coho{O}(\mb{T,T,T})=\epsilon$ (all others are $1$).
Therefore USp$(4)_2$ possesses an $\mathcal{H}^3(\mathbb{Z}_2^{\bf T}, \mathbb{Z}_2)$ 
symmetry-localization anomaly.

\section{Sufficient conditions for the presence of 
$\mathcal{H}^3_{[\rho]}(\mathbb{Z}_2^{\bf T},\mathcal{A})$ anomaly}
\label{constraints}

\subsection{Overview}

The above discussion of the $\mathcal{H}^3$ symmetry-localization anomaly requires detailed
knowledge of the $F$ and $R$ symbols of the theory in order to determine whether any symmetry $G$
and map $[\rho]$ possesses the anomaly. However obtaining the $F$ and $R$ symbols given the modular
data (the $S$ matrix and the topological twists) of a TQFT is often a computationally prohibitive problem. 
Below we will discuss some simple conditions that must be satisfied for a theory to be free of the 
$\mathcal{H}^3_{[\rho]}(\mathbb{Z}_2^{\bf T},\mathcal{A})$ anomaly. 

First we define the quantities:
\begin{align}
\mathcal{Z}(\mathbb{RP}^4) &= \sum_{\{ a | a = \,^{\bf T}a \}} S_{0a} \theta_a \eta_a^{\bf T} ,
\nonumber \\
M_a &= \sum_{\{ x | x = \,^{\bf T}x \}} S_{ax} \eta_x^{\bf T} , 
\end{align}
whose significance will be described in the subsequent sections. 

The conditions that must be satisfied are as follows:
\begin{enumerate}
\item $\mathcal{Z}(\mathbb{RP}^4) = \pm 1$.
\item $M_a$ is a non-negative integer for all $a$. 
\item $\theta_a = \pm 1$ and $\theta_a$ is independent of $a$, for all $a$ such that $M_a > 0$.
\end{enumerate}
Given the modular $S$-matrix, the topological twists, and an action of ${\bf T}$ that permutes the anyons,
there must be a choice of $\{ \eta_a^{\bf T} \}$ such that conditions (1)-(3) are satisfied. If not, the
theory possesses an $\mathcal{H}^3_{[\rho]}(\mathbb{Z}_2^{\bf T},\mathcal{A})$ anomaly. 

For the case of reflection symmetry ${\bf R}$, we replace $^{\bf T}a$ with $\,^{\bf R}\bar{a}$, and $\eta_a^{\bf T}$ with
$\eta_a^{\bf R}$ in the above formulae:
\begin{align}
\mathcal{Z}(\mathbb{RP}^4) &= \sum_{\{ a | a = \,^{\bf R}\bar{a} \}} S_{0a} \theta_a \eta_a^{\bf R} ,
\nonumber \\
M_a &= \sum_{\{ x | x = \,^{\bf R}\bar{x} \}} S_{ax} \eta_x^{\bf R} .
\end{align}

In the subsequent sections we will discuss these conditions and their origin in detail. It will be convenient to phrase
the discussion in terms of spatial reflection symmetry ${\bf R}$, and then to obtain the results for time reversal ${\bf T}$
by replacing ${\bf R}$ with ${\bf CT}$ (where ${\bf C}: a \rightarrow \bar{a}$ is topological charge conjugation) 
and $\{\eta_a^{\bf R}\}$ with $\{\eta^{\bf T}_a\}$. 

In Sec. \ref{conflictingEta}, we also provide an additional diagnostic by considering the theory obtained by condensing 
certain bosons in the TQFT of interest, where the inconsistency arises by finding conflicting constraints on $\{\eta_a^{\bf T}\}$. 

In order to connect the constraints above to $\mathcal{H}^3_{[\rho]}(G, \mathcal{A})$ 
symmetry-localization anomalies, we need to make an important assumption that we discuss below.
Let us consider a (2+1)D topological phase with symmetry $G$ whose symmetry
action $[\rho]$ is free of the $\mathcal{H}^3_{[\rho]}(G, \mathcal{A})$ symmetry-localization anomaly. 
Depending on the symmetry fractionalization class, which recall is classified by $\mathcal{H}^2_{[\rho]}(G, \mathcal{A})$,
the system may possess an SPT (t 'Hooft) anomaly, in the sense that the fractionalization class can only occur if the (2+1)D system
is realized at the surface of a (3+1)D SPT state. The (3+1)D SPT cancels the anomaly from the symmetry fractionalization
of the (2+1)D surface. In this way, a given (2+1)D SET determines a (3+1)D SPT state. If the (2+1)D theory has no anomaly
at all, then the bulk (3+1)D system is a trivial SPT. 

In the following, we will show that the choice of symmetry action $[\rho]$ alone can be 
enough to preclude the (2+1)D system from existing at the surface of a (3+1)D SPT state. 
In the example of USp$(4)_2$ CS theory, we find that the existence of the $\mathcal{H}^3_{[\rho]}(G, \mathcal{A})$ 
anomaly is accompanied by the impossibility of the theory to be consistent at the surface of any
(3+1)D $\mathbb{Z}_2^{\bf T}$ SPT state. We expect that this is a general
phenomenon: the impossibility of the theory, with a specified action of $[\rho]$, to exist at the surface of a (3+1)D SPT state
with symmetry group $G$ signals the existence of the $\mathcal{H}^3_{[\rho]}(G, \mathcal{A})$ 
anomaly. In principle, for space-time reflection symmetries we have not ruled out the possibility that 
the failure of the (2+1)D system to exist at the surface of a (3+1)D SPT state could signal an additional 
anomaly associated with $[\rho]$, which is independent of the $\mathcal{H}^3_{[\rho]}(G, \mathcal{A})$ obstruction. 
However this appears to be unlikely, given that this is known not to be the case for unitary internal 
symmetries.\cite{barkeshli2014SDG,ENO2009} 

\subsection{Condition (1): (3+1)D Path integral on $\mathbb{RP}^4$}

In the case of $\mathbb{Z}_2^{\bf R}$ (or  $\mathbb{Z}_2^{\bf T}$) symmetry, bosonic SPTs in (3+1)D 
have a $\mathbb{Z}_2 \times \mathbb{Z}_2$ classification. The 4 distinct SPT states can be 
distinguished by the value of their topological path integrals on $\mathbb{RP}^4$ and $\mathbb{CP}^2$:\cite{kapustin2014}
\begin{align}
\label{4dcond}
\mathcal{Z}(\mathbb{RP}^4) &= \pm 1 ,
\nonumber \\
\mathcal{Z}(\mathbb{CP}^2) &= \pm 1. 
\end{align}

As mentioned above, a given (2+1)D SET determines a (3+1)D SPT state. In Ref. \onlinecite{barkeshli2016tr}, 
it was shown that if the symmetry of the system is $\mathbb{Z}_2^{\bf R}$, then one can 
compute the path integral on $\mathbb{RP}^4$ entirely in terms of the properties of the (2+1)D theory 
through the following formula:
\begin{align}
\mathcal{Z}(\mathbb{RP}^4) &= \sum_{\{a | a = \,^{\bf R}\bar{a}\}} S_{0a} \theta_a \eta_a^{\bf R} .
\end{align}
Here, the sum is over all anyons that are invariant under the action of reflection, ${\bf R}$, and 
topological charge conjugation. 

Similarly, one can compute the path integral on $\mathbb{CP}^2$:
\begin{align}
\mathcal{Z}(\mathbb{CP}^2) &= \frac{1}{\mathcal{D}} \sum_a d_a^2 \theta_a = e^{2\pi i c/8 } ,
\end{align}
where $c$ is the chiral central charge of the UMTC. 

Therefore, for any (2+1)D topological phase that admits a $\mathbb{Z}_2^{\bf R}$ symmetry with a
given action $[\rho]$, there must be a choice of $\{\eta_a^{\bf R}\}$ such that (\ref{4dcond}) is 
satisfied. Failure to sastify (\ref{4dcond}) for any choice of $\{ \eta_a^{\bf R} \}$ implies that the symmetry 
action $[\rho]$ cannot be realized in a way that allows the system to exist at the surface of a (3+1)D SPT state. 

For the case of the $\mr{USp}(4)_2$ CS theory, performing the computation for $\mathbb{RP}^4$ (using
time-reversal $\mathbb{Z}_2^{\bf T}$ instead of reflection ${\bf Z}_2^{\bf R}$ as the example), we find:
\begin{align}
\mathcal{Z}(\mathbb{RP}^4) = \frac{1}{\mathcal{D}} (1 + \eta_{\epsilon}^{\bf T}). 
\end{align}
We can see that if $\eta_{\epsilon}^{\bf T} = -1$, then $\mathcal{Z}(\mathbb{RP}^4) = 0$. 
If $\eta_{\epsilon}^{\bf T} = 1$, then $\mathcal{Z}(\mathbb{RP}^4) = \frac{2}{\mathcal{D}} = \frac{1}{\sqrt{5}}$. 
We see that in both cases, the bulk (3+1)D system cannot be a $\mathbb{Z}_2^{\bf T}$ SPT state.

\subsection{Condition (2): constraints from $\mathbb{RP}^2 \times D^2$}

Let us suppose that we are given a $\mathbb{Z}_2^{\bf R}$ SET state that can exist at the surface of a (3+1)D 
SPT state. Consider the path integral for the (3+1)D theory on $\mathbb{RP}^2 \times D^2$, where $D^p$ denotes
the $p$-dimensional disk. We choose boundary conditions on the boundary, $\mathbb{RP}^2 \times S^1$, such 
that there is a Wilson loop of $a$ encircling the $S^1$ at a particular point on the $\mathbb{RP}^2$. 
In Ref. \onlinecite{barkeshli2016tr}, this path integral, denoted as $\mathcal{Z}(\mathbb{RP}^2 \times D^2)[l_a] $, 
was computed to be
\begin{align}
\mathcal{Z}(\mathbb{RP}^2 \times D^2)[l_a] \equiv M_a = \sum_{\{x | x = \,^{\bf R}\bar{x}\}} S_{ax} \eta_x^{\bf R} . 
\end{align}

Below we derive the following non-trivial constraint, that $M_a$ must be a non-negative integer:
\begin{align}
\label{Maint}
M_a \in \mathbb{Z}_{\geq 0} .
\end{align}

To derive (\ref{Maint}), let us first consider the case where the bulk (3+1)D theory is a trivial SPT, 
i.e. when $\mathcal{Z}(\mathbb{RP}^4) = \mathcal{Z}(\mathbb{CP}^2) = 1$. In this case,
all path integrals of the (3+1)D theory depend solely on the (2+1)D boundary. Thus
$M_a$ can be intepreted to be solely a property of the boundary of $\mathbb{RP}^2 \times D^2$:
\begin{align}
M_a = \mathcal{Z}_{2+1;a}(\mathbb{RP}^2 \times S^1), 
\end{align}
where the subscript (2+1) emphasizes that this is purely a property of the (2+1)D theory. 
This, in turn, corresponds to the dimension of the Hilbert space of the (2+1)D theory on 
$\mathbb{RP}^2$, with a puncture labelled by $a$. By definition the dimension of a 
Hilbert space must be a non-negative integer; thus $M_a$ must be a non-negative integer. 

Now let us consider the case when the bulk (3+1)D theory is a non-trivial SPT, i.e. 
when either $\mathcal{Z}(\mathbb{RP}^4) = -1$ or  
$\mathcal{Z}(\mathbb{CP}^2) = -1$. In this case we argue for (\ref{Maint}) in two steps.
We first prove that $2 M_a \in \mathbb{Z}_{\geq 0}$, after which we prove that $M_a M_b \in \mathbb{Z}$.
Together these imply (\ref{Maint}). 

To prove that $2 M_a \in \mathbb{Z}_{\geq 0}$, we argue as follows. 
Let us refer to the bulk (3+1)D theory of interest as $A$, which can be obtained from 
the UMTC $\mathcal{C}$, which below we denote as $\mathcal{C}_A$. Let us now consider 
a second bulk (3+1)D theory, denoted $B$, which can be obtained from a second UMTC $\mathcal{C}_B$, and which satisfies 
$\mathcal{Z}^B(\mathbb{RP}^4) = \mathcal{Z}^A(\mathbb{RP}^4)$, and
$\mathcal{Z}^B(\mathbb{CP}^2) = \mathcal{Z}^A(\mathbb{CP}^2)$. 

Next, we consider the combined theory, denoted $AB$, such that
\begin{align}
\mathcal{Z}^{AB} (M^4) = Z^A(M^4) Z^B(M^4) ,
\end{align}
where $M^4$ is any closed $4$-manifold. Now consider
\begin{align}
M^{AB}_{(a,b)} = M^A_a M^B_b ,
\end{align}
where $M^{AB}_{(a,b)} = \mathcal{Z}^{AB}(\mathbb{RP}^2 \times D^2)[l_{(a,b)}]$,
$a \in \mathcal{C}_A$, and $b \in \mathcal{C}_B$. By our previous argument, we have
$M^{AB}_{(a,b)} \in \mathbb{Z}_{\geq 0}$. 

When $\mathcal{Z}^B(\mathbb{RP}^4) = -1$ and $\mathcal{Z}^B(\mathbb{CP}^2) = 1$,
we can take $\mathcal{C}_B$ to be the $\mathbb{Z}_2$ toric code model, 
which has four particles, $\{1, e, m, \psi\}$, with $\psi = e \times m$ the fermion. Furthermore, we
consider the case where $\eta_e^{\bf R} = \eta_m^{\bf R} = -1$. For this theory, 
$M^B_b = 2 \delta_{b \psi}$. Since we also have $M^{AB}_{(a,b)} \in \mathbb{Z}_{\geq 0}$, it follows that 
$2 M^A_a \in \mathbb{Z}_{\geq 0}$. 

When $\mathcal{Z}^B(\mathbb{RP}^4) = -1$ and $\mathcal{Z}^B(\mathbb{CP}^2) = -1$,
we can take $\mathcal{C}_B$ to be the three-fermion theory, i.e. SO$(8)_1$ CS theory.
This is an Abelian theory with four particle types, $\{1, f_1, f_2, f_3 \}$, where $f_i$ are 
fermions ($\theta_{f_i} = -1$), with $S_{f_i, f_j} = -1/2$ for $i \neq j$. 
For this theory, $M^B_b = 2 \delta_{b 1}$, and therefore again $2 M^A_a \in \mathbb{Z}_{\geq 0}$
in this case as well. 

Finally, when $\mathcal{Z}^B(\mathbb{RP}^4) = 1$ and  $\mathcal{Z}^B(\mathbb{CP}^2) = -1$,
we can take $\mathcal{C}_B$ to consist of the three-fermion model, with
$\eta_{f_1}^{\bf R} = -1$ and $\eta_{f_2}^{\bf R} = 1$. Then we get $M^B_b = 2 \delta_{b f_2}$,
which again proves that $2 M^A_a \in \mathbb{Z}_{\geq 0}$. 

Next, we must prove that $M_a M_b \in \mathbb{Z}$. This follows from the fact that 
\begin{align}
\label{Mabform}
\mathcal{Z}(\mathbb{RP}^2 \times S^1 \times I)[l_a \cup l_b] &= M_a M_b ,
\end{align}
where the boundary conditions (denoted in the square brackets) are such that there is a loop of $a$ encircling the $S^1$ 
at $\mathbb{RP}^2 \times \{0\}$ and a loop of $b$ encircling the $S^1$ at $\mathbb{RP}^2 \times \{1\}$.
Here we take the interval $I = [0,1]$, so that $\{0\}$ and $\{1\}$ are the boundaries of $I$. 
$\mathcal{Z}(\mathbb{RP}^2 \times S^1 \times I)[l_a \cup l_b] $ can be interpreted as the dimension 
of the Hilbert space of the (3+1)D system on $\mathbb{RP}^2 \times I$
with two anyons $a$ and $b$ on the two ends of $I$. Since this is the dimension of a Hilbert space,
$M_a M_b$ must be a non-negative integer.  

It remains to prove (\ref{Mabform}). We can demonstrate (\ref{Mabform}) as follows, using ideas from Ref. \onlinecite{walker2006}.\footnote{We thank
K. Walker for helpful discussions regarding this point.} We refer the reader to Ref. \onlinecite{barkeshli2016tr} for a 
detailed discussion of these computations written for physicists. 
First we note that the (3+1)D path integrals on closed manifolds are all bordism invariant when constructed from UMTCs. 
Therefore,
\begin{align}
\label{RP2S2}
\mathcal{Z}(\mathbb{RP}^2 \times S^2) = 1,
\end{align}
because $\mathbb{RP}^2 \times S^2$ is bordant to the empty manifold, as it is the boundary of 
$\mathbb{RP}^2 \times D^3$. Next, we use the fact that the Hilbert space of the (3+1)D theory
on any closed $3$-manifold is one-dimensional, which follows from the fact that this is a bulk (3+1)D 
SPT, with no intrinsic topological order. Therefore, using the gluing formula for topological path integrals,\cite{walker2006,barkeshli2016tr}
\begin{widetext}
\begin{align}
\mathcal{Z}(\mathbb{RP}^2 \times S^2) = 
\frac{ \mathcal{Z}(\mathbb{RP}^2 \times D^2)[l_a] \mathcal{Z}(\mathbb{RP}^2 \times S^1 \times I)[l_a \cup l_b] \mathcal{Z}(\mathbb{RP}^2 \times D^2)[l_b]}{\langle l_a | l_a \rangle_{\mathcal{V}(\mathbb{RP}^2 \times S^1)} \langle l_b |l_b \rangle_{\mathcal{V}(\mathbb{RP}^2 \times S^1) }} .
\end{align}
\end{widetext}
Here, we consider obtaining $\mathbb{RP}^2 \times S^2$ by gluing $\mathbb{RP}^2 \times S^1 \times I$ to two copies of 
$\mathbb{RP}^2 \times D^2$, each one along $\mathbb{RP}^2 \times S^1$. Thus the inner product is in the Hilbert space
$\mathcal{V}(\mathbb{RP}^2 \times S^1)$, as indicated by the subscript. 
Using (\ref{RP2S2}), moving the denominator to the LHS, and using the definition of the inner product in terms of 
the path integral\cite{walker2006,barkeshli2016tr} we obtain
\begin{align}
\mathcal{Z}&(\mathbb{RP}^2 \times S^1 \times I)[l_a \cup l_{\bar a}] \mathcal{Z}(\mathbb{RP}^2 \times S^1 \times I)[l_b \cup l_{\bar b}]
\nonumber \\
&= M_a M_b \mathcal{Z}(\mathbb{RP}^2 \times S^1 \times I)[l_a \cup l_b] . 
\end{align}
Together with the identity $M_a = M_{\bar a}$, this implies (\ref{Mabform}).

\subsection{Condition (3): Topological twists and $M_a$}

The third non-trivial constraint summarized above is
\begin{align}
\label{Matwist}
\theta_a &= \pm 1, \;\; \text{ and independent of $a$, for } M_a > 0. 
\end{align}
In the special case that $M_1 > 0 $, this implies that $\theta_a = 1$ for all $a$ such that $M_a > 0$. 

One way to derive (\ref{Matwist}) is to use the following identity, shown in Ref. \onlinecite{barkeshli2016tr}:
\begin{align}
\label{productIdentity}
\frac{\sum_a \theta_a M_a^2}{\sum_a M_a^2} = \mathcal{Z}(\mathbb{RP}^4) \mathcal{Z}(\mathbb{CP}^2) . 
\end{align}
Since $M_a^2$ is a non-negative integer, and $\mathcal{Z}(\mathbb{RP}^4) \mathcal{Z}(\mathbb{CP}^2) = \pm 1$,
the condition (\ref{Matwist}) follows trivially. 

To better understand the origin of (\ref{Matwist}), below we will prove it through a different approach, which does not make use of the 
identity in Eq. (\ref{productIdentity}). Let us again consider the path integral of the (3+1)D system,
$\mathcal{Z}(\mathbb{RP}^2 \times S^1 \times I)[l_a \cup l_b] = M_a M_b$.
As in the previous section, the boundary conditions denoted in square brackets consist
of a Wilson loop of $a$ along the $S^1$ direction on the first $\mathbb{RP}^2 \times S^1$, and a Wilson loop of $b$
along the $S^1$ direction on the other $\mathbb{RP}^2\times S^1$. 

Now, we can consider the following procedure. If $M_a M_b > 0$, 
we can consider a state $|\Psi_{a, b}\rangle$ from the Hilbert space 
on $\mathbb{RP}^2 \times I$, which contains a puncture of $a$ and $b$ on the top and bottom surfaces. 
Next, we can cut out a tube that encircles $a$ on the top $\mathbb{RP}^2$ 
and $b$ on the bottom surface, rotate the tube by $2\pi$, and glue it back in. This leads to an operation
$\mathcal{T}$ on the state $|\Psi_{a, b}\rangle$ which corresponds to a Dehn twist on the top surface, and the oppositely oriented
Dehn twist on the bottom surface. Thus we obtain:
\begin{align}
\mathcal{T} |\Psi_{a b} \rangle = \theta_a \theta_{b}^* |\Psi_{a b} \rangle.
\end{align}
Since the Dehn twist around the cross-cap on $\mathbb{RP}^2$ is isotopic to the identity (see e.g. Ref. \onlinecite{barkeshli2016tr} 
for a discussion of this ), it is a trivial operation and therefore we must have that $\mathcal{T} |\Psi_{ab}\rangle = |\Psi_{ab}\rangle$. 
In particular, we must have
\begin{align}
\theta_a \theta_{b}^* = 1. 
\end{align}
This implies that $\theta_a = \theta_b$ whenever $M_a M_b > 0 $; that is, that $\theta_a$ is independent
of $a$ for all $a$ such that $M_a > 0$. 

Next, we can consider taking $b$ through the cross-cap on the bottom $\mathbb{RP}^2$, obtaining 
$\,^{\bf R}b$ and yielding the state $|\Psi_{a, \,^{\bf R}b} \rangle$. The Dehn twist procedure now gives
\begin{align}
\mathcal{T} |\Psi_{a \,^{\bf R}b} \rangle = \theta_a \theta_{\,^{\bf R}b}^* |\Psi_{a \,^{\bf R}b} \rangle
\end{align}
Again, since the Dehn twist operation is trivial, this gives
\begin{align}
\theta_a = \theta_{\,^{\bf R}b} .
\end{align}
In particular, if we take $b = a$, we find:
\begin{align}
\theta_a = \theta_{\,^{\bf R}a} . 
\end{align}
Since we also know that the reflection action must satisfy
\begin{align}
\theta_a = \theta_{\,^{\bf R}a}^* , 
\end{align}
we find that $\theta_a$ must be real. This proves (\ref{Matwist}). 

Examining the example of $\mr{USp}(4)_2$ (again using time-reversal $\mathbb{Z}_2^{\bf T}$ as the example), we find:
\begin{align}
M_a = S_{a1} + S_{a\epsilon} \eta_{\epsilon}^{\bf T} .
\end{align}
We see that $M_a$ will all be integer only if we set $\eta_{\epsilon}^{\bf T} = -1$.
However, this then implies that $M_{\psi_+} = M_{\psi_-} > 0$. But 
$\theta_{\psi_+} = \theta_{\psi_-}^* = i \neq \pm 1$. We thus conclude again that 
$\mr{USp}(4)_2$ cannot be a $\mathbb{Z}_2^{\bf T}$ time-reversal invariant topologial phase that 
exists on the surface of a (3+1)D $\mathbb{Z}_2^{\bf T}$ SPT. 

\subsection{Conflicting constraints on symmetry fractionalization quantum numbers}
\label{conflictingEta}

Here we point out that another symptom of the $\mathcal{H}^3_{[\rho]}(\mathbb{Z}_2^{\bf T}, \mathcal{A})$  
symmetry localization anomaly arises in terms of conflicting constraints on the time-reversal symmetry
fractionalization patterns.

On the one hand, Ref. \onlinecite{barkeshli2014SDG} derived a set of constraints for $\eta^{\bf T}_a$, which 
showed that
\begin{align}
\eta_a^{\bf T} = \theta_a , \;\;\; \text{ if $N_{c \,^{\bf T}c}^a$ is odd, for any $c$.  }
\end{align}

In the case of $\mr{USp}(4)_2$, we have $\psi_+ \times \psi_- = \epsilon + \cdots$, 
and $\,^{\bf T} \psi_+ = \psi_-$. Therefore, we must have
\begin{align}
\eta_\epsilon^{\bf T} = 1 . 
\end{align}

On the other hand, we show below that by a different argument, we must have
\begin{align}
\eta_\epsilon^{\bf T} = -1 . 
\label{eqn:epsilonT2}
\end{align}
It is the incompatibility of the above two results that signals the presence of an $\mathcal{H}^3$ anomaly
for the $\mathbb{Z}_2^{\bf T}$ symmetry. 

To see how Eq. \eqref{eqn:epsilonT2} arises, we observe that since $\epsilon$ is an Abelian boson, we can 
condense it, leading us to a new topological phase, which in this case corresponds 
to $\mr{SU}(5)_1$ CS theory. To be more explicit, we label the anyons in $\mr{SU}(5)_1$ by $[j]$, 
with $j=0,\dots, 4 \text{ (mod 5)}$. They form a $\mathbb{Z}_5$ group under fusion. 
The topological twists are $\theta_j=e^{\frac{4\pi i }{5}j^2}$. This theory has a charge conjugation symmetry ${\bf C}: [j]\leftrightarrow [-j]$. 

In addition, we observe that SU(5)$_1$ CS theory has a $\Z_4^{\bf T}$ symmetry, where 
\begin{align}
	{\bf T}: [j] \rightarrow [2j] ,
	\label{eqn:TinSU5}
\end{align}
so that ${\bf T}^2 = \mb{C}$, and $\mb{C}^2 = 1$.  Notice that similar to USp(4)$_2$,  the chiral central
charge $c = 4$ implying that the theory can only be time-reversal invariant at the surface of a (3+1)D theory. 

The inverse process of condensing an Abelian $\Z_2$ boson in USp(4)$_2$ is to gauge the unitary $\Z_2$ symmetry ${\mb C}$ in SU(5)$_1$. 
The general procedure for gauging is discussed in Ref. \onlinecite{barkeshli2014SDG}. To see how gauging ${\mb C}$ in SU$(5)_1$ yields
USp$(4)_2$, let us enumerate the anyons in the gauged theory.
First we enlarge the theory to include extrinsic $\mathbb{Z}_2$ symmetry defects, 
denoted by $\psi$. $\psi$ satisfies the following fusion rules:
\begin{align}
	\psi\times\psi &=[0]+[1]+[2]+[3]+[4], 
\nonumber \\
\psi\times [a] &=\psi.
	\label{}
\end{align}
We then need to project the whole theory to the $\mathbb{Z}_2$-invariant subspace, which in the mathematics literature is referred to as ``equivariantization.''
This amounts to (a) reorganizing objects into orbits under the symmetry, and (b) including symmetry charges. In the present case, the nontrivial anyons in $\mathrm{SU}(5)_1$ form two orbits: $\{[1],[4]\}$ and $\{[2],[3]\}$, which correspond to $\phi_1$ and $\phi_2$. Both the identity and the defect split into 
two, carrying opposite $\mathbb{Z}_2$ charges, which are $1, \epsilon$ for the identity and $\psi_+, \psi_-$ for $\psi$.

Since ${\bf T}^2$ is gauged, and since $\epsilon$ is the $\mathbb{Z}_2$ gauge charge, we must have that
\begin{align}
\eta_\epsilon^{\bf T} = -1 ,
\end{align}
because $\eta_{\epsilon}^{\bf T}$ is precisely the local ${\bf T}^2$ value of $\epsilon$. The anyon permutation given in
 Eq. \eqref{eqn:TinSU5} becomes $\phi_1\leftrightarrow \phi_2$ in the gauged theory.

Another possible way to see that $\eta_{\epsilon}^{\bf T} = -1$ follows from the general constraints of 
Eq. (\ref{kappaEtaEq}). While we do not pursue this analysis in detail here, we note that a similar analysis was performed 
in the case of D$(\mathbb{S}_3)$ in the Appendix of Ref. \onlinecite{barkeshli2016tr}.
Here D$(\mathbb{S}_3)$ refers to the quantum double of $\mathbb{S}_3$, the permutation
group on three elements.

\subsection{Non-anomalous cousins}
\label{sec:checktheory}

We have glossed over an important detail of the gauging procedure, upon which we now elaborate.  As discussed in 
Ref. \onlinecite{barkeshli2014SDG}, once the $\mathbb{Z}_2$ symmetry fractionalization class is 
chosen, the remaining $\mathbb{Z}_2$ symmetry enriched phases are related to each other by elements of
$\H^3(\mathbb{Z}_2, \U) = \mathbb{Z}_2$. The difference between these two can be thought of as stacking a 
$\mathbb{Z}_2$ (2+1)D SPT before gauging (in high energy field theory language, this corresponds to adding a Dijkgraaf-Witten\cite{dijkgraaf1990}
term for the $\mathbb{Z}_2$ gauge field). The SPT phase does not change either the bulk anyons or the chiral central charge. The only effect is that the 
topological twist factors of the $\mathbb{Z}_2$ gauge fluxes $\psi_\pm$ are modified by a factor of $i$. Namely, we 
will have $\theta_{\psi_+}=1, \theta_{\psi_-}=-1$ instead. 

\begin{table}
	\begin{tabular}{c|c|c|c|c|c|c}
          \hline 
          $a$ & $1$ & $\epsilon$ & $\phi_1$ & $\phi_2$ & $\psi_+$ & $\psi_-$ \\
          \hline
		  $\theta_a$ & $1$ & $1$ & $e^{\frac{4\pi i}{5}}$ & $e^{-\frac{4\pi i}{5}}$ & $1$ & $-1$ \\
          \hline
          $d_a$ & $1$ & $1$ & $2$ & $2$ & $\sqrt{5}$ & $\sqrt{5}$ \\
          \hline
          $\,^{\bf T}a$ & $1$ & $\epsilon$ & $\phi_2$ & $\phi_1$ & $\psi_+$ & $\psi_-$ \\
        \end{tabular}

\caption{Anyon types, topological spins, and quantum dimensions for USp(4)$^\vee_2$.
\label{USpchecktable}
}
\end{table}

This implies that USp(4)$_2$ has a partner theory with the same fusion rules, where $\psi_\pm$ are invariant under the action of
time-reversal (see Table \ref{USpchecktable}). Let us refer to this theory as $\mr{USp}(4)^{\vee}_2$. We can readily compute for this theory that
\begin{align}
\mathcal{Z}(\mathbb{RP}^4) = \frac{1}{\sqrt{20}} ( 1 + \eta_\epsilon^{\bf T} + \sqrt{5} \eta_{\psi_+}^{\bf T} - \sqrt{5} \eta_{\psi_-}^{\bf T}).
\end{align}
Using $\eta_\epsilon^{\bf T} = -1$ and $\eta_{\psi_-}^{\bf T} = \eta_{\epsilon}^{\bf T} \eta_{\psi_+}^{\bf T}$, we obtain:
\begin{align}
\mathcal{Z}(\mathbb{RP}^4) = \eta_{\psi_+}^{\bf T} = \pm 1. 
\end{align}

Since $\psi_+$ and $\psi_-$ are both time-reversal invariant, we no longer have the constraint
$\eta_\epsilon^{\bf T} = \theta_\epsilon$. Therefore, USp(4)$_2^\vee$ does not lead to conflicting
constraints on $\eta_{\epsilon}^{\bf T}$. We conclude that USp(4)$_2^\vee$ does not possess
the $\mathcal{H}^3_{[\rho]}(\mathbb{Z}_2^{\bf T}, \mathcal{A})$ anomaly. 
Depending on the value of $\eta_{\psi_+}^{\bf T}$, it does have the SPT ('t Hooft) anomalies associated with
$\mathcal{Z}(\mathbb{RP}^4) $ and $\mathcal{Z}(\mathbb{CP}^2)$, which indicate which
(3+1)D $\mathbb{Z}_2^{\bf T}$ SPT state hosts it at the surface. In particular, if we see 
$\eta_{\psi_+}^\mb{T}=1$, the USp(4)$_2^\vee$ theory only has the $\mathcal{Z}(\mathbb{CP}^2)$ 
anomaly, which matches the SPT (t' Hooft) anomaly in SU(5)$_1$.

Another way of stating the relation between USp(4)$_2$ and USp(4)$_2^\vee$ is as follows. 
We can stack a double semion state\cite{freedman2004} on top of USp(4)$_2^\vee$. Note that the double semion state
is described by U$(1)_2 \times $U$(1)_{-2}$ CS theory. We denote the four anyons in the double semion state by 
$1, s, s', b=s\times s'$, where $s$ ($s'$) is a semion(anti-semion). The resulting theory has $24$ particles. 
In particular, it has a boson $(b, \epsilon)$, consisting of the boson from the double semion state and the 
$\epsilon$ from the USp(4)$_2^\vee$ state. If we condense the composite $(b, \epsilon)$, 
the following particles remain deconfined after the condensation:
\begin{equation}
	\begin{gathered}
	1\sim (b, \epsilon), (1,\epsilon)\sim (b,1), (1, \phi_i)\sim (b, \phi_i)\\
	(s, \psi_+)\sim (s', \psi_-), (s, \psi_-)\sim (s', \psi_+),
	\end{gathered}
	\label{}
\end{equation}
where $\sim$ here means differing by the condensed particle. We can readily see that the resulting theory is USp(4)$_2$.

However, this condensation process must break the $\Z_2^\mb{T}$ symmetry. 
This is because the $b$ boson must have $\eta_b^\mb{T}=\theta_b=1$, because $b=s\times s'$ and 
$s'={}^\mb{T}s$ due to the opposite topological spins. Therefore, the bound state $(b, \epsilon)$ also has 
$\eta_{(b,\epsilon)}^{\bf T} = -1$, i.e. it carries a local Kramers degeneracy. Condensing $(b, \epsilon)$ 
must therefore break time-reversal symmetry.

To summarize, we have found a cousin theory $\mr{USp}(4)_2^\vee$ of $\mr{USp}(4)_2$ which has at most t'Hooft anomaly.

\section{Three Resolutions}
\label{resolutions}

Above we saw that the $\mathbb{Z}_2^{\bf T}$ symmetry in $\mr{USp}(4)_2$  possesses an $\mathcal{H}^3$ 
anomaly, implying that the symmetry action cannot be consistently localized to the quasiparticles. In the 
following we discuss three possible resolutions of this anomaly.

\subsection{Enlarging the symmetry from $\Z_2^\mb{T}$ to $\mathbb{Z}_4^{\bf T}$}
\label{Z4T}

One resolution of the $\mathbb{Z}_2^{\bf T}$ symmetry localization anomaly is that the theory 
actually does not have a $\mathbb{Z}_2^{\bf T}$ symmetry, but rather the true symmetry is $\mathbb{Z}_4^{\bf T}$. 

Mathematically, one can show straightforwardly that if the symmetry group is enlarged to $\mathbb{Z}_4^\mb{T}$, 
the obstruction class we found earlier (naturally embedded into the larger symmetry group) becomes trivial. 
Here instead we will present a physical argument, explicitly constructing the $\mathrm{USp}(4)_2$ theory in a system with $\mathbb{Z}_4^\mb{T}$ time-reversal symmetry.  

Let us start from a system of bosons where each boson, $\phi$, has ${\bf T}^2=-1$ (e.g. spin-$1/2$ bosons).
Thus globally $\mb{T}^2=(-1)^{N_\phi}$ on the whole system, where $N_\phi$ is the number of bosons. Microscopically
the system therefore possesses a $\Z_4^{\bf T}$ symmetry. Pairs of bosons
in this system thus locally have ${\bf T}^2 = 1$. Let us suppose that the paired bosons realize a 
topological phase described at long wavelengths by two decoupled theories, consisting of the double
semion state and the $\mr{USp}(4)_2^{\vee}$ state. Now consider the composite $(b, \epsilon)$ 
(recall $b$ is the topologically non-trivial boson of the double semion state). Depending on whether
we attach the fundamental ${\bf T}^2=-1$ boson $\phi$, the composite $(b, \epsilon)$ can be 
a Kramers singlet, so that its condensation no longer breaks any symmetry. This way we have a realization of $\mr{USp}(4)_2$ with 
$\Z_4^\mb{T}$ symmetry (up to the chiral central charge anomaly, which can be cancelled by a bulk (3+1)D SPT). 

We notice that similar resolutions apply to unitary symmetry groups as well. For example, the $\mathcal{H}^3$ obstruction 
for a $\mathbb{Z}_2$ symmetry in the $\mathbb{D}_{16}$ gauge theory can be avoided if the symmetry group is 
actually $\mathbb{Z}_4$.\cite{fidkowski2015}

\subsection{$\mr{USp}(2N)_N$ CS theory as a fermionic (spin) theory with SPT (t 'Hooft) anomaly}

In Ref. \onlinecite{aharony2016}, it was proposed that $\mr{USp}(2N)_N$ CS theory is time-reversal 
invariant as a fermionic theory. Here we consider the case where ${\bf T}^2 = -1$ on the electron 
creation/annihilation operators, so that globally ${\bf T}^2 = (-1)^{N_f}$ on the state of the system, 
with $(-1)^{N_f}$ being the fermion parity of the system. We show that
$\mr{USp}(4)_2$ possesses a time-reversal anomaly that can be cancelled by a bulk (3+1)D electronic
time-reversal-invariant topological superconductor in class DIII. In other words, the theory no longer
has any $\mathcal{H}^3$ anomaly, but it does have an SPT anomaly. This may look similar to the 
$\Z_4^\mb{T}$ symmetry we discussed in the previous section, however we cannot use the same 
argument to resolve the anomaly due to the fermionic statistics.

Recall that $\mr{USp}(4)_2$ has a cousin $\mr{USp}(4)_2^\vee$ that is free of the 
$\mathcal{H}^3(\mathbb{Z}_2^{\bf T}, \mathcal{A})$ anomaly. As discussed earlier in Sec. \ref{sec:checktheory}, the 
two theories are related by stacking with a double semion state and condensing certain bosonic quasiparticles. 
To preserve the time-reversal symmetry, the boson $b$ in the double semion has to have $\eta_b^\mb{T}=-1$, 
which is impossible in bosonic systems in two dimensions. However, if the double semion theory is viewed
as a state arising from microscopic degrees of freedom that contain fermions (i.e. as a spin theory), 
then time-reversal symmetry can be implemented differently:
\begin{align}
	\mb{T}: &s\leftrightarrow s \times f, 
\nonumber \\
&s'\leftrightarrow s' \times f.
	\label{}
\end{align}
Here $f$ denotes the local fermion, which is transparent under braiding with the anyons of the system.  
With this action of ${\bf T}$, time-reversal symmetry changes the local fermion parity on the semion and 
anti-semion. As shown in Ref. \onlinecite{metlitski2014}, if an anyon $a$ transforms as $a \rightarrow a \times f$ 
under ${\bf T}$, then (i) $f$ must have ${\bf T}^2=-1$, i.e. the fermions are Kramers doublets; and 
(ii) $a$ has a well-defined ``${\bf T}^2$'' value which can now be $\pm i$ and which will also be denoted by 
$\eta_a^\mb{T}$. Such anyons are said to carry a ``Majorana Kramers doublet''. However, unlike in bosonic systems, 
for two anyons $a$ and $b$ both having Majorana Kramers doublets, and $c$ with $N_{ab}^c=1$, one can show that
\begin{equation}
	\eta_c^\mb{T} = -\eta_a^\mb{T} \eta_b^{\mb{T}}.
	\label{}
\end{equation}

In the double semion state above, since both $s$ and $s'$ carry Majorana Kramers doublets we have 
$\eta_b^\mb{T}=-\eta_s^\mb{T}\eta_{s'}^\mb{T}$. Therefore, if $\eta_s^{\bf T} =i, \eta_{s'}^{\bf T} =-i$, we have 
$\eta_b^\mb{T}=-1$. This is precisely the surface topological order of a $\nu=4$ class DIII topological 
superconductor ($\nu=1$ is the root phase, having a single Majorana cone on the surface without any interactions). 

With this anomalous fermionic double semion theory, we can condense $(b,\epsilon)$ 
in $\mr{USp}(4)_2^\vee$ stacked with the double semion state,
obtaining a ${\bf T}$-invariant $\mr{USp}(4)_2$ theory 
with ${\bf T}^2 = (-1)^{N_f}$. This provides a surface topological order for the $\nu = 4$ class DIII
topological superconductor. Under ${\bf T}$, the anyons are transformed according to
\begin{equation}
	\mb{T}: \phi_1\leftrightarrow \phi_2, \psi_\pm \leftrightarrow \psi_\pm f.
	\label{}
\end{equation}

We note that this conclusion is consistent with the use of the fermionic anomaly indicator formula
discussed in Ref. \onlinecite{wang2016,tachikawa2016b}.

\subsection{Pseudo-2D realization at the surface of (3+1)D SETs}
\label{pseudo2D}

While strictly speaking $\mathbb{Z}_2^{\bf T}$ cannot be a symmetry of theories 
with an $\mathcal{H}^3$ anomaly, below we will demonstrate a precise sense in which 
these theories can admit a $\mathbb{Z}_2^{\bf T}$ symmetry, provided they are ``pseudo-realized'' at the
surface of a (3+1)D topologically ordered bulk state. From the discussion of Sec. \ref{sec:checktheory}, 
we see that it is sufficient to show that the anomalous double semion state (where $\eta_b^{\bf T} = -1$) 
can be ``pseudo-realized.''

\subsubsection{Pseudo-realizing double semion with $\eta_b^{\bf T} = -1$ on the surface of a (3+1)D SET} 

As discussed in the preceding sections, in (2+1)D a double semion state with $\mathbb{Z}_2^{\bf T}$ time-reversal symmetry
must satisfy $\eta_b^{\bf T} = \theta_b =1$ because $N_{s \,^{\bf T}s}^b = 1$. 

Nevertheless, below we will show how one can realize a state that is closely related (but strictly speaking not identical) 
to the double semion state with $\eta_b^{\bf T} = -1$ on the surface of a bulk (3+1)D system with long-range entanglement.  
To understand this, it is helpful to consider spatial reflection symmetry, ${\bf R}$, instead of time-reversal symmetry. 
In the TQFT, we replace ${\bf T}$ by ${\bf C R}$. Since charge conjugation ${\bf C}$ acts trivially here, the action of ${\bf T}$
can just be replaced with the action of ${\bf R}$. 

\subsubsection{$\mathbb{Z}_2$ gauge theory on the bulk mirror plane}

In Ref. \onlinecite{song2016}, it was shown how (3+1)D reflection invariant SPTs (with ${\bf R}^2 = \mathds{1}$), can
be understood by restricting attention to the mirror plane. On the mirror plane, the reflection symmetry 
acts like an on-site $\mathbb{Z}_2$ symmetry. Thus, to understand (3+1)D SPTs with $\mathbb{Z}_2^{\bf R}$ reflection symmetry, 
one is led to considering the possible existence of non-trivial (2+1)D SPT states with on-site (internal) $\mathbb{Z}_2$ symmetry
existing on the mirror plane. 

Here, we instead consider a (2+1)D SET on the mirror plane. Specifically, on the mirror plane we consider a 
$\mathbb{Z}_2$ toric code state, which is described by a dynamical $\mathbb{Z}_2$ (untwisted) gauge theory.
Furthermore, we consider a global $\mathbb{Z}_2^{\bf R}$ reflection symmetry, which acts as an on-site $\mathbb{Z}_2$ 
symmetry on the mirror plane. The anyons in $\mathbb{Z}_2$ gauge theory are denoted by $\{1, e, m, \psi\}$ where $e$ ($m$) is the $\mathbb{Z}_2$ gauge charge (flux). 
We further consider the case where the $e$ and $m$ particles of the $\mathbb{Z}_2$ gauge theory both carry
fractional $\mathbb{Z}_2$ charge under the global $\mathbb{Z}_2$ symmetry. 

\begin{figure}[t!]
	 \centering
	 \includegraphics[width=3in]{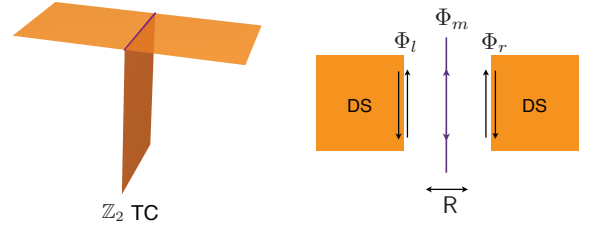}
	 \caption{Illustration of the construction of the anomalous double semion state on the surface of a $\mathbb{Z}_2$ toric code on the mirror plane. Left panel: 3D setup, where a $\mathbb{Z}_2$ gauge theory harbors the mirror plane. Right panel: top view of the surface.}
	 \label{fig:3dmirror}
 \end{figure}

Next, we consider a (2+1)D surface of the (3+1)D bulk system. At a given time-slice, the (2+1)D surface
forms a plane which is perpendicular to the mirror plane of the (3+1)D bulk on which the $\mathbb{Z}_2$ gauge theory lives (see Fig. \ref{fig:3dmirror}). 
We consider the case where the surface theory is a double semion state. Below we will demonstrate that this surface theory
can have $\eta_b^{\bf R} = -1$. 

To demonstrate this explicitly, we imagine cutting the system along the mirror axis on the surface, so that we have 
three subsystems: the double semion edge states on the left and right of the mirror plane, and the $\mathbb{Z}_2$ gauge 
theory on the mirror plane. We need to show that at the $(1+1)$ dimensional intersection between the surface and the bulk mirror plane, 
 the three pairs of edge modes can be fused together in a way which preserves the global symmetry and which is gapped everywhere. 
Note that with the on-site $\mathbb{Z}_2$ symmetry, the $\mathbb{Z}_2$ gauge theory has gapless edge modes, 
since a gapped edge has to correspond to either $e$ or $m$ condensation, which necessarily breaks the symmetry because both of them carry half charge. 

To this end, we see that the $(1+1)$D theory at the junction consists of the edge theories of the three subsystems. Each of them admits a description as a non-chiral Luttinger liquid, and altogether can be compactly written as
\begin{equation}
	\mathcal{L}=\frac{1}{4\pi}\partial_t\Phi^\mathsf{T} K \partial_x \Phi - \dots
	\label{}
\end{equation}
here $\Phi=(\phi_{1l}, \phi_{2l}, \phi_{1r}, \phi_{2r}, \phi_{1m}, \phi_{2m})^\mathsf{T}$, where $\phi_{l/r}$ refer to the edge bosonic fields 
of the double semion states on the left/right, and $\phi_m$ refer to the edge fields of the $\mathbb{Z}_2$ gauge theory on the mirror plane. The $K$-matrix reads
\begin{align}
K = \left( \begin{matrix} 2 \sigma^z & 0 &0 \\
0 & 2 \sigma^z &0  \\
0 & 0 & 2 \sigma^x \end{matrix} \right)
\end{align}
The $\pm 2\sigma^z$ parts describe the contribution from the double semion state on either side of the interface.
The $2 \sigma^x$ part describes the contribution from the $\mathbb{Z}_2$ gauge theory on the mirror plane.
We adopt the convention that the $e$ particle at the mirror plane corresponds to the operator $e^{i \phi_{1m}}$, and the $m$ particle
corresponds to the operator $e^{i \phi_{2m}}$.  The $\mathbb{Z}_2^{\bf R}$ symmetry acts in the 
following way on these fields:
\begin{equation}
	\begin{split}
		\phi_{1m} &\rightarrow \phi_{1m} +  \frac{\pi}{2}
\nonumber \\
\phi_{2m} &\rightarrow \phi_{2m} + \frac{\pi}{2}
\nonumber \\
\phi_{1l} &\leftrightarrow \phi_{1r} 
\nonumber \\
\phi_{2l} &\leftrightarrow \phi_{2r} 
	\end{split}
	\label{eqn:R on fields}
\end{equation}
The first two transformations on $\phi_{1m}$ and $\phi_{2m}$ encode the fact that the $e$ and $m$ particles carry
half $\mathbb{Z}_2$ charge. The boson operator on one side of the interface is $b = e^{i (\phi_{1l} + \phi_{2l})}$,
and on the other side of the interface is $b = e^{i (\phi_{1r} + \phi_{2r})}$. 

The idea in this construction will be that the boson on the surface will be the $\mathbb{Z}_2$ gauge charge on the mirror plane. 
Therefore we consider the following gapping terms:
\begin{equation}
	\begin{split}
		\delta \mathcal{L} = & - u\cos 2( \phi_{1l} + \phi_{2l} - \phi_{1m})\\
		&+ u\cos 2 (\phi_{1r} + \phi_{2r} - \phi_{1m}) 
\\
&-v\cos 2 (\phi_{1l} - \phi_{2l} + \phi_{1r} - \phi_{2r} - 2\phi_{2m} )
\end{split}
	\label{}
\end{equation}
The gapping terms preserve the reflection symmetry defined in Eq. \eqref{eqn:R on fields}. The first two 
terms show that the combination $b \times e$ is condensed at the junction, which implies that
at the junction, $b$ can continue into the mirror plane as the $e$ particle, which is the $\mathbb{Z}_2$ gauge charge. 

The above gapping terms can be written as 
\begin{align}
\delta\mathcal{L} =	\sum_{a} t_a \cos (\bm{\Lambda}_{a}^\mathsf{T} K \Phi), 
\end{align}
where the integer vectors $\bm{\Lambda}_a$ determine the gapping terms. In the present case, we have
\begin{align}
\bm{\Lambda}_1 &= (1, 1, 0, 0, -1, 0)
\nonumber \\
\bm{\Lambda}_2 &= (0, 0, 1, 1, -1, 0)
\nonumber \\
\bm{\Lambda}_3 &= (1, -1, 1, -1, 0, -1)
\end{align}
and $t_1 = -t_2 = u$, $t_3 = v$. One can see that these vectors are null vectors for the $K$-matrix, as they satisfy
\begin{align}
	\bm{\Lambda}_a^\mathsf{T} K \bm{\Lambda}_b = 0
\end{align}
for all $a, b = 1,2, 3$. We should also check that the global $\mathbb{Z}_2$ symmetry is not spontaneously broken in the ground state. 
Using the criteria derived in Ref. \onlinecite{LevinPRB2012}, we can in fact show that the gapping terms lead 
to a unique gapped ground state, thus excluding the possibility of spontaneous symmetry breaking.

Importantly, observe that the operator $e^{i (\phi_{1l} - \phi_{2l}) + i (\phi_{1r} -\phi_{2r})}$ creates a pair of $b$ bosons on either side
of the mirror plane, in a mirror-symmetric way. Due to the third gapping term above, we see that since 
$e^{i (\phi_{1l} - \phi_{2l} + \phi_{1r} -\phi_{2r} + 2 \phi_{2m})}$ is condensed, then $e^{i (\phi_{1l} - \phi_{2l}) + i (\phi_{1r} -\phi_{2r})}$
can be replaced by $e^{2i \phi_{2m}}$. Since under the global $\mathbb{Z}_2^{\bf R}$ symmetry $\phi_{2m} \rightarrow \phi_{2m} + \pi/2$, 
we see that this operator goes to minus itself, which is precisely the definition of the reflection eigenvalue $\eta_b^{\bf R} = -1$. 

\subsubsection{$\mathbb{Z}_2$ gauge theory in the bulk (3+1)D system}

\begin{figure}[t!]
	\centering
	\includegraphics[width=2.0in]{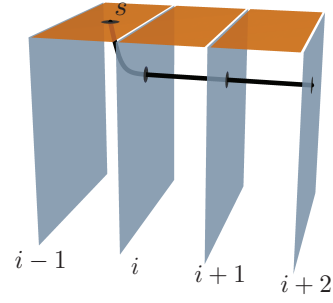}
	\caption{Illustration of the layer construction. Each vertical layer denotes a 
$\mathbb{Z}_2$ toric code, and neighboring layers are coupled such that pairs of 
$e$ particles are condensed. The surface consists of a double semion state, joined to the
vertical layers at each (1+1)D intersection by condensing $b \times e$. }
	\label{fig:layer}
\end{figure}

In the previous section we showed how the double semion state with $\eta_b^{\bf R} = -1$ can be realized at the surface
of a bulk (3+1)D system with a certain $\mathbb{Z}_2$ gauge theory on the bulk mirror plane. Here we will briefly point out 
that with this starting point, one can then consider a layer construction, where we stack (2+1)D  $\mathbb{Z}_2$ toric code
states on planes parallel to the mirror plane (see Fig. \ref{fig:layer}). We condense pairs of $e$ particles from neighboring 
planes, so that the $e$ particle can propagate in three dimensions. The other deconfined excitations are strings of $m$ 
particles from each layer. This way we get a bulk (3+1)D $\mathbb{Z}_2$ gauge theory, with a global $\mathbb{Z}_2^{\bf R}$ reflection symmetry. 

At the interface of the bulk planes with the (2+1)D surface we condense pairs of $e$ and $b$ particles together, so that the
$b$ particle can just propagate from the surface into the bulk. Because of the condensation of 
$b\times e$ at each intersection, if a semion/anti-semion were to propagate on the surface, 
each time it passes through an intersection it has to bind with a $m$ particle from the layer. 
In other words, due to the mutual statistics of the particles, the condensation of $b \times e$ 
confines the semions/anti-semions $s$ and $s'$ to the end points of the $m$-strings.

We refer to the above as a ``pseudo-realization'' of the double semion state with 
$\eta_b^{\bf R} = -1$ at the surface of a (3+1)D $\mathbb{Z}_2$ gauge theory. The reason 
we call it a ``pseudo-realization'' is because strictly speaking, the double semion is not confined to live on the surface
of the (3+1)D system anymore. The boson $b$ can propagate into the bulk, while the semions at the surface are bound to the end-points of $\mathbb{Z}_2$ flux
strings in the bulk. In this case, since the endpoints of the flux strings do not have a well-defined topological spin, it is no longer meaningful
to associate them with semions. 

We thus find that while the double semion state with $\eta_b^{\bf R} = -1$ is strictly speaking not allowed, it can be
``pseudo-realized'' at the surface of a bulk (3+1)D SET with global $\mathbb{Z}_2^{\bf R}$ reflection symmetry. 

\subsection{Pseudo-realizing $\mr{USp}(4)_2$}

Combining the above results, we can now conclude the following. $\mr{USp}(4)_2$ possesses an $\mathcal{H}^3(\mathbb{Z}_2^{\bf T},\mathcal{A})$ 
time-reversal anomaly. It has a cousin $\mr{USp}(4)_2^{\vee}$ which does not possess this $\mathcal{H}^3$ time-reversal anomaly. One
can obtain $\mr{USp}(4)_2$ from $\mr{USp}(4)_2^{\vee}$ by stacking a double semion on the latter, and condensing the combination of
$\epsilon \times b$. If we demand that $\eta_b^{\bf T} = -1$, then the bound state will have $\eta_{b \epsilon}^{\bf T} = 1$, and therefore
condensing it will not break ${\bf T}$. However such a double semion state with $\mathbb{Z}_2^{\bf T}$ symmetry can 
only be pseudo-realized at the surface of a bulk $(3+1)$D that contains a dynamical $\mathbb{Z}_2$ gauge field. 
Therefore $\mr{USp}(4)_2$ can be pseudo-realized with $\mathbb{Z}_2^{\bf T}$ symmetry at the surface of a 
bulk $(3+1)D$ $\mathbb{Z}_2$ gauge theory.

{{
\section{More examples of theories with $\mathcal{H}^3$ anomalies}
\label{examples}

We have seen that the $\mathcal{H}^3_{[\rho]}(\mathbb{Z}_2^{\bf T}, \mathcal{A})$ anomaly in USp$(4)_2$ CS theory 
is intimately associated with the $\mathbb{Z}_4^{\bf T}$ symmetry in SU$(5)_1$ CS theory. Below we will 
see that this is a general phenomenon. We provide an infinite series of TQFTs with 
$\mathcal{H}^3_{[\rho]}(\mathbb{Z}_2^{\bf T}, \mathcal{A})$ anomalies, obtained by gauging 
${\bf T}^2$ in a theory with $\mathbb{Z}_4^{\bf T}$ symmetry. 

The essential point is that when we gauge ${\bf T}^2$, we can consider adding a Dijkgraaf-Witten (DW) term 
for the gauge field associated with the ${\bf T}^2$ symmetry. Physically this corresponds to stacking
a $\mathbb{Z}_2$ SPT (associated with ${\bf T}^2$) together with the theory with the $\mathbb{Z}_4^{\bf T}$ 
symmetry. This combined system no longer possesses $\mathbb{Z}_4^{\bf T}$ symmetry, because a $\mathbb{Z}_2$ SPT 
is incompatible with $\mathbb{Z}_4^{\bf T}$ symmetry if the $\mathbb{Z}_2$ SPT is to be protected by ${\bf T}^2$.
We will see that the theory obtained by gauging ${\bf T}^2$ under these circumstances thus leads us to another theory
that contains the $\mathcal{H}^3$ anomaly. 

\subsection{Gauging ${\bf T}^2$ in $\mr{USp}(4)_2$ CS theory}
\label{UspInfSeries}

Given that $\mr{USp}(4)_2$ supports a $\mathbb{Z}_4^{\bf T}$ symmetry, we can
study the time-reversal symmetry properties of the theories obtained by gauging 
the global unitary $\mathbb{Z}_2$ symmetry ${\bf T}^2$.\cite{barkeshli2014SDG} 

The symmetry fractionalization associated with the $\mathbb{Z}_2$ global symmetry is classified by
$\mathcal{H}^2(\mathbb{Z}_2, \mathbb{Z}_2) = \mathbb{Z}_2$. In one case, none of the particles
of $\mr{USp}(4)_2$ carry fractional quantum numbers; naively gauging the $\mathbb{Z}_2$ gives rise to 
two decoupled theories: $\mr{USp}(4)_2 \times D(\mathbb{Z}_2)$, where $D(\mathbb{Z}_2)$ refers to a 
$\mathbb{Z}_2$ discrete gauge theory. Depending on whether there is a Dijkgraaf-Witten (DW) term
in the effective action for the $\mathbb{Z}_2$ gauge field, the $D(\mathbb{Z}_2)$ factor will correspond 
to a toric code or double semion theory. One then runs into the contradiction that since the $\mathbb{Z}_2$ 
gauge theory is decoupled from $\mr{USp}(4)_2$, the theory continues to have the same 
$\mathcal{H}^3_{[\rho]}(\mathbb{Z}_2^{\bf T}, \mathcal{A})$ anomaly, which is impossible if 
we start from an anomaly-free theory. 

The resolution is that we have not been careful enough in gauging. The correct theory obtained from gauging can be most easily 
obtained from the physical construction in Sec. \ref{Z4T}. There we observed that $\mr{USp}(4)_2$ with $\mathbb{Z}_4^{\bf T}$ symmetry
can be obtained as follows. We start with a theory of physical bosons $\phi$, which are the local degrees of freedom of the theory,
and which have ${\bf T}^2 = -1$. Then we imagine that pairs of bosons form a topological state described by $\mr{USp}(4)_2^\vee \times \text{DS}$.
Next we condense the particle $b\epsilon$ combined with the local boson $\phi$. Since $b \epsilon$ has $\mb{T}^2 = -1$, combining it
with $\phi$ yields a particle which is a Kramers singlet, which can then be condensed without breaking the $\mathbb{Z}_4^{\bf T}$ symmetry. 

Thus to gauge $\mb{T}^2$ in $\mr{USp}(4)_2$, we first start from $\mr{USp}(4)_2^\vee\times \text{DS}$, then we gauge 
the $\mb{T}^2$ symmetry and subsequently condense the particle that corresponds to the fusion of $b \epsilon$ and the $\mathbb{Z}_2$ charge
of the ${\bf T}^2$ gauge field. 

Since we are considering the case with no DW term and trivial symmetry fractionalization classes, the result 
for gauging $\bf{T}^2$ in $\mr{USp}(4)_2^\vee \times \text{DS}$ is $\mr{USp}(4)_2^\vee\times \mr{DS} \times \mr{D}(\Z_2)$. 
Here $\mr{D}(\Z_2)$ refers to an untwisted $\mathbb{Z}_2$ gauge theory. Denote the four particles in $\mr{D}(\Z_2)$ as 
$\{1,e, m,\psi=e\times m\}$ where $e$ is interpreted as the gauge charge and therefore has $\mb{T}^2=-1$. 
Thus we now condense $\epsilon b e$. After condensation, we can identify an Abelian subsector
\begin{equation}
	\begin{gathered}
	1, b\sim \epsilon e, sm \sim s'\psi \epsilon, s'm\sim s\psi\epsilon\\
	\epsilon, \epsilon b\sim e, \epsilon sm \sim s'\psi, \epsilon s'm\sim s\psi
	\end{gathered}
	\label{}
\end{equation}
The first line is nothing but a DS theory. We can further identify another sector
\begin{equation}
	1, \epsilon, \psi_+ m, \psi_- m, \phi_1, \phi_2, 
	\label{}
\end{equation}
which is closed under fusion and can again be identified as $\mr{USp}(4)_2^\vee$. Thus the resulting theory is 
$\mr{USp}(4)_2^\vee\times \mr{DS}$, and $\mb{T}$ acts diagonally on the two sectors. This theory is free of any 
$\mathcal{H}^3$ anomaly, as it should be. If we further add a DW term in gauging the $\mb{T}^2$ symmetry, we would 
obtain a theory with $\H^3$ anomaly again, which turns out to be $\mr{USp}(4)_2\times \mr{D}(\Z_2)$.

In the case where the $\mr{USp}(4)_2$ particles do carry fractional quantum numbers (corresponding to the non-trivial $\mathbb{Z}_2$
fractionalization class), we can obtain the gauged theory as follows. 

In the physical construction of Sec. \ref{Z4T}, we let the $s$ and $s'$ particles in the DS sector both have fractional quantum numbers, i.e.
\begin{equation}
	\eta_s(\mb{T}^2,\mb{T}^2)=\eta_{s'}(\mb{T}^2,\mb{T}^2)=-1.
	\label{}
\end{equation}
One can easily see that after condensing $b\epsilon$ together with a physical boson $\phi$, we do get a $\mr{USp}(4)_2$ where 
$\psi_\pm$ have fractional quantum numbers under $\mb{T}^2$. 

The gauged theory can be obtained as follows: first we add a gauge flux $\sigma$. Due to the fractionalization of 
$\mb{T}^2$, the flux satisfies the fusion rule $\sigma^2=b$. The topological twist of this flux can be chosen to be 
either $\theta_\sigma=1$ or $\theta_\sigma=i$, depending on whether a DW term is included or not. Without loss of generality we also set
\begin{equation}
	M_{s, \sigma}=i, M_{s',\sigma}=-i,
	\label{}
\end{equation}
where recall $M_{ab}$ is the braiding phase between $a$ and $b$, defined in Eq. \ref{Mabdef}.
Further we also need to add a bosonic $\mb{T}^2$ gauge charge $e$, such that $s^2=s'^2=e$. Notice that we still define $b=ss'$.

When $\theta_{\sigma}=1$, we can identify the gauged theory as $\mr{USp}(4)_2^\vee \times \mr{D}(\Z_4)$. 
The $\Z_4$ gauge charge, which we label as $(1,0)$, is identified with $\sigma$. The $\Z_4$ gauge flux $(0,1)$
can be identified with $s' \sigma e=s\sigma b$. The ${\bf T}^2$ gauge charge $e$ becomes 
$(2,2)$.  For reference, the topological twist of an anyon $(a,b)$ in D$(\Z_4)$ is
\begin{equation}
	\theta_{(a,b)}=i^{ab}.
	\label{}
\end{equation}

Under time reversal symmetry $\mb{T}$ (which now satisfies $\mb{T}^2=\mb{1}$), we find that 
$e$ and $\sigma$ are invariant, while $s \sigma b \rightarrow s \sigma e=\overline{s\sigma b}$. 
We have $\eta_\sigma^{\bf T} = 1$, $\eta_b^{\bf T} = 1$, $\eta_e^{\bf T} = -1$, and $\eta_{be}^{\bf T} = -1$,
which form a consistent set of time-reversal symmetry fractionalization quantum numbers. Therefore, as expected,
the case where $\theta_{\sigma} = 1$, which corresponds to no added DW term, does not have a $\mathcal{H}^3$ anomaly. 

When $\theta_{\sigma}=i$, we again define $(1,0)\equiv \sigma, (0,1)\equiv s\sigma b$, with 
$\theta_{(0,1)}=\theta_{(1,0)}=i$. A general anyon $(a,b)$ in this theory has topological twist
\begin{equation}
	\theta_{(a,b)}=i^{a^2+b^2-ab}.
	\label{}
\end{equation}
Let us denote it by $\mr{D}'(\Z_4$). 

Now we determine the time-reversal transformation in this case. Using the consistency of braiding, we can uniquely fix
\begin{align}
	{}^\mb{T}\sigma &=\overline{\sigma}e.
\nonumber \\
	{}^\mb{T} s \sigma b &=s \sigma.
\end{align}

This theory $\mr{D}'(\Z_4$) has a $\H^3$ anomaly: on the one hand, $\eta_{e}^\mb{T}=-1$ due to the fact that $e$ is a $\mb{T}^2$ 
gauge charge of the original theory. However, on the other hand we have $\eta_{be}^{\bf T} = \theta_{be} = 1$, because $be = \,^{\bf T}\sigma \times \sigma$. 
Since $\eta_b^{\bf T} = 1$, it follows that $\eta_{e}^{\bf T} = \eta_{be}^{\bf T} \eta_b^{\bf T} = 1$, which is a contradiction. 

We now start from $\mr{USp}(4)_2^\vee\times \mr{D}(\Z_4)$ or $\mr{USp}(4)_2^\vee\times \mr{D}'(\Z_4)$ depending on whether $\theta_\sigma=1$ or $i$, and condense $\epsilon b e\equiv \epsilon \times (0,2)$.  Notice that while $\mr{D}(\Z_4)$ and $\mr{D}'(\Z_4)$ generally have different braiding structures, the braiding of $(0,2)$ with other anyons are the same:
\begin{equation}
	M_{(a,b),(0,2)}=(-1)^a.
	\label{}
\end{equation}
Therefore we can treat the condensation uniformly for both cases.
The resulting gauged theory contains $24$ particles, with the spins and quantum dimensions shown in Table \ref{tab:USpgaugeZ2}. 

\begin{table}
	\centering
	\begin{tabular}{|c|c|c|c|}
		\hline
		Label & $d$ & $\theta$ & Remark\\
		\hline
		$(a,b)$ & $1$ & $\theta_{(a,b)}$ & $a\in \{0,2\}, b\in \{0,1,2,3\}$\\
		\hline
		$\psi_+\times (a,b)$ & $\sqrt{5}$ & $\theta_{(a,b)}$ & $a\in\{1,3\}, b \in \{0,1,2,3\}$\\
		\hline
		$\phi_1\times (a,b)$ & $2$ & $e^{\frac{4\pi i}{5}}\theta_{(a,b)}$ & $a\in \{0,2\}, b \in \{0,1\}$\\
		\hline
		$\phi_2\times (a,b)$ & $2$ & $e^{-\frac{4\pi i}{5}}\theta_{(a,b)}$ &  $a\in \{0,2\}, b \in \{0,1\}$\\
		\hline
	\end{tabular}
	\caption{Particle types, quantum dimensions, and topological twists for the theory obtained by gauging a unitary global $\mathbb{Z}_2$ symmetry
in $\mr{USp}(4)_2$. We take the case where the symmetry fractionalization
class is non-trivial. There are thus two remaining distinct choices, $\theta_{\sigma} = 1, i$, encoded in the two expressions for $\theta_{(a,b)}$}. 
	\label{tab:USpgaugeZ2}
\end{table}

When $\theta_\sigma = i$, from Table \ref{tab:USpgaugeZ2} we see that all of the anyons must be permuted by ${\bf T}$ because
of their complex twists, except for $(0,0), (2,0), (0,2)$ and $(2,2)$, which are invariant under ${\bf T}$. This implies that 
\begin{equation}
	\mathcal{Z}(\mathbb{RP}^4) = \frac{1}{2 \sqrt{20}} (1 + \eta_{(2,0)} ^{\bf T} + \eta_{(0,2)} ^{\bf T} + \eta_{(2,2)}^{\bf T}) \neq \pm 1. 
	\label{}
\end{equation}
We thus conclude that this theory possesses a $\mathcal{H}^3_{[\rho]}(\mathbb{Z}_2^{\bf T}, \mathcal{A})$ anomaly. 

To summarize, we have found so far that USp$(4)_2$ CS theory possesses an $\mathcal{H}^3(\mathbb{Z}_2^{\bf T}, \mathcal{A})$ anomaly.
As discussed above, one resolution to this is that the symmetry can be taken to be $\mathbb{Z}_4^{\bf T}$. This allows us to consider the theories obtained
by gauging ${\bf T}^2$, which leads us to new theories that also possess $\mathcal{H}^3(\mathbb{Z}_2^{\bf T}, \mathcal{A})$ anomalies,
for example as summarized in Table \ref{tab:USpgaugeZ2}. Again this means that the true symmetry of the new gauged theory
is $\mathbb{Z}_4^{\bf T}$. Thus we can again consider gauging ${\bf T^2}$. This procedure can
continue indefinitely, yielding an infinite family of theories with $\mathcal{H}^3(\mathbb{Z}_2^{\bf T}, \mathcal{A})$ anomalies. 
 }}

\subsection{$\mr{SO}(4)_4$ CS theory}

Another example of a theory with an $\mathcal{H}^3(\mathbb{Z}_2^{\bf T}, \mathcal{A})$ anomaly
is $\mr{SO}(4)_4$ CS theory. 

$\mr{SO}(4)_4$ CS theory can be obtained from $\mr{SU}(2)_4 \times \mr{SU}(2)_4$ CS theory by 
a condensation process as follows. The 5 particle types of SU$(2)_4$ CS theory can be organized
in terms of SU$(2)$ representations: $0, 1/2, 1, 3/2, 2$, with the spin $2$ particle being the Abelian anyon.
Thus the particle types of $\mr{SU}(2)_4 \times \mr{SU}(2)_4$ can be written as $(a,b)$, with $a,b = 0, 1/2, 1, 3/2, 2$. 
$\mr{SO}(4)_4$ CS theory can be obtained from $\mr{SU}(2)_4 \times \mr{SU}(2)_4$ CS theory by condensing
the spin $(2,2)$ Abelian boson.\cite{moore1989} This leads to a theory with $8$ types of particles, summarized in Table \ref{SO44table}.

\begin{table}
	\begin{tabular}{c|c|c|c|c|c|c|c|c}
          \hline 
          $a$ & $1$ & $(1/2, 1/2)$ & $(0,1)$ & $(1,0)$ & $(1/2,3/2)$ & $(1,1)_+$ & $(1,1)_-$ & $(0,2)$  \\
          \hline
		  $\theta_a$ & $1$ & $i$ & $e^{\frac{2\pi i}{3}}$ & $e^{\frac{2\pi i}{3}}$ & $-i$ & $e^{-\frac{2\pi i }{3}}$ & $e^{-\frac{2\pi i}{ 3}}$ & $1$ \\
          \hline
          $d_a$ & $1$ & $3$ & $2$ & $2$ & $3$ & $2$ & $2$ & $1$ \\
          \hline
        \end{tabular}
\caption{Anyon types, topological spins, and quantum dimensions for SO(4)$_4$.
\label{SO44table}
}
\end{table}

$\mr{SO}(4)_4$ has an Abelian particle $\epsilon = (1,1)$. Condensing $\epsilon$ takes us to a new theory, 
$\mr{SU}(3)_1 \times \mr{SU}(3)_1$ CS theory. $\mr{SU}(3)_1$ CS theory is an Abelian theory with 3 particle types. 
Thus we can label the anyons of $\mr{SU}(3)_1 \times \mr{SU}(3)_1$ 
as $(a,b)$, for $a,b = 1,\cdots 3 \text{ (mod 3)}$. This theory has a $\mathbb{Z}_4^{\bf T}$ symmetry associated with the 
following transformation on the anyons:
\begin{align}
{\bf T}: (a,b) \rightarrow (2a+b, a+b)
\end{align}
Note that the above transformation induces the following action:
$(1,0) \rightarrow (2,1) \rightarrow (2,0) \rightarrow 1,2) \rightarrow (1,0)$, and $(0,1) \rightarrow (1,1) \rightarrow (0,2)
\rightarrow (2,2) \rightarrow (0,1)$. It is easy to verify that ${\bf T}^2 = {\bf C}$, and therefore ${\bf T}$ generates a
$\mathbb{Z}_4^{\bf T}$ symmetry. We note that it is possible to obtain another $\mathbb{Z}_4^{\bf T}$ symmetry 
by defining ${\bf T}' = {\bf L} {\bf T} {\bf L}$, where ${\bf L}$ is the $\mathbb{Z}_2$ layer-exchange symmetry,
${\bf L} : (a,b) \rightarrow (b,a)$. 

This theory has all of the same essential features that appeared in our preceding
analysis of $\mr{USp}(4)_2$ CS theory. The possible actions of ${\bf T}$ preclude 
the constraints discussed in Sec. \ref{constraints} from being satisfied, in a similar manner to that of $\mr{USp}(4)_2$
CS theory. For example, 
\begin{align}
\mathcal{Z}(\mathbb{RP}^4) = \frac{1}{6}(1 + \eta_{(0,2)}^{\bf T}) \neq \pm 1. 
\end{align}

Moreover, it is straightforward to check that all of the same resolutions as discussed for 
$\mr{USp}(4)_2$ apply in this case as well. 

As in the $\mr{USp}(4)_2$ case, we can now consider enlarging the symmetry to $\mathbb{Z}_4^{\bf T}$, which does
not possess the $\mathcal{H}^3$ anomaly, and subsequently gauge the ${\bf T}^2$ symmetry. Through this process,
one can again generate an infinite series of TQFTs with $\mathcal{H}^3$ anomalies, with now $\mr{SU}(3)_1 \times \mr{SU}(3)_1$
CS theory being the ``root'' phase, instead of $\mr{SU}(5)_1$ CS theory. 

\subsection{Infinite family of root phases}
\label{KmatrixExSec}

Given the understanding developed in the preceding sections of when $\mathcal{H}^3$ anomalies arise, we can
now provide an infinite family of ``root phases'' with $\mathbb{Z}_4^{\bf T}$ symmetry which, upon gauging 
${\bf T}^2$, each lead to a series of theories with $\mathcal{H}^3$ anomalies.  Let us consider a two-component Abelian 
$\mr{U}(1) \times \mr{U}(1)$ CS theory:
\begin{equation}
	\mathcal{L}_\mr{CS}=\frac{1}{4\pi}\sum_{I,J=1}^2\varepsilon^{\mu\nu\lambda}a_{I\mu} K_{IJ} \partial_\nu a_{J\lambda} + \cdots,
	\label{}
\end{equation}
with the $K$-matrix
\begin{align}
\label{KmatrixEx}
K = \left( \begin{matrix}
m & n \\
n & -m 
\end{matrix} \right).
\end{align}
where $n, m$ are integers, and $m$ is even to describe a bosonic (non-spin) theory. 
The quasiparticles of this theory are described by 2-component integer vectors $\vec{l}$. 

This theory has a time-reversal symmetry, whose action on the gauge fields is given by 
\begin{align}
{\bf T}:
\begin{pmatrix}
	a_1\\
	a_2
\end{pmatrix}
\rightarrow
\begin{pmatrix}
a_2\\
-a_1
\end{pmatrix}.
\end{align}
It is clear that
\begin{align}
	{\bf T}^2 &= -\mathds{1} ,
\end{align}
which takes all quasiparticles $\vec{l} \rightarrow - \vec{l}$. Therefore, 
${\bf T}^2$ is a $\mathbb{Z}_2$ charge-conjugation symmetry, while
${\bf T}$ generates a $\mathbb{Z}_4^{\bf T}$ symmetry. 

It is important to know that the $\mathbb{Z}_4^{\bf T}$ symmetry of the $K$-matrix in Eq. (\ref{KmatrixEx}) 
does not possess any anomaly. In Appendix \ref{coupledWireAppendix}, we give an explicit microscopic construction 
of this phase with a $\mathbb{Z}_4^{\bf R}$ reflection symmetry in (2+1)D, to explicitly demonstrate the absence of any anomaly. 

We note that the case where $(m,n) = (0,3)$ gives rise to $\mathbb{Z}_3$ gauge theory. 
Gauging the unitary $\mathbb{Z}_2$ particle-hole symmetry gives rise to $\mathbb{S}_3$
gauge theory (the permutation group on three elements), which contains 8 particles.\cite{barkeshli2014SDG} It is 
a close cousin of the $\mr{SO}(4)_4$ example, but with chiral central charge $c = 0$.

Similarly, the case where $(m,n) = (2,1)$ gives rise to a $\mathbb{Z}_5$ anyon theory (i.e. the fusion rules
form a $\mathbb{Z}_5$ group. This is a close cousin of $\text{SU}(5)_1$ CS theory).  
Gauging the unitary $\mathbb{Z}_2$ particle-hole symmetry gives rise to a theory
with $6$ particles, which is closely related to $\mr{USp}(4)_2$, but with $c= 0$. 

In fact we can define a $\mathbb{Z}_4^\mb{T}$ symmetry for an even more general class of anyon models 
denoted as $\mathbb{Z}_N^{(p)}$, for $N\equiv 1\, (\text{mod }4)$. $\mathbb{Z}_N^{(p)}$ contains Abelian anyons
with $\mathbb{Z}_N$ fusion rules and with $R$ symbols $R^{[a][b]}_{[a+b]} = e^{2\pi i p a b/N}$. 
The $\mb{T}$ symmetry acts on an anyon $[j]$ as
\begin{equation}
	\mb{T}: [j]\rightarrow [tj],
	\label{}
\end{equation}
such that $\theta_{[tj]}=\theta_{[j]}^*$, which implies $t^2\equiv -1\,(\text{mod }N)$. Since $N$ is odd, 
$t$ has to be even in order to satisfy this condition. As a result, we must have 
$N\equiv 1\, (\text{mod }4)$ as a necessary condition. $\mb{T}^2$ is
automatically the charge-conjugation symmetry.

\subsubsection{Gauging the unitary $\mathbb{Z}_2$ symmetry}

Let us now consider gauging the unitary $\mathbb{Z}_2$ charge conjugation symmetry associated with
${\bf T}^2$ in the theories described above. We will show that this leads to a theory with an 
$\mathcal{H}^3_{[\rho]}(\mathbb{Z}_2^{\bf T}, \mathcal{A})$ anomaly. 
As discussed in Ref. \onlinecite{barkeshli2014SDG}, there are in principle
two distinct ways to gauge the $\mathbb{Z}_2$ symmetry, corresponding to a choice
of group element in $\mathcal{H}^3(\mathbb{Z}_2, \mathrm{U}(1)) = \mathbb{Z}_2$. 
This is associated with whether we include a Dijkgraaf-Witten
term for the $\mathbb{Z}_2$ gauge field in the effective Lagrangian description. 
(Note that for this theory this is the only choice that needs to be made in the gauging process, 
because the symmetry fractionalization class $\mathcal{H}^2_{[\rho]}(\mathbb{Z}_2^{\bf T}, \mathcal{A})$ is trivial
in this case\cite{barkeshli2014SDG}). 

Here we mainly focus on the case $\text{gcd}(m,n) = 1$. In this case, the $K$-matrix defined above 
defines an Abelian theory with $\mathbb{Z}_{m^2 + n^2}$ fusion rules.\footnote{See, e.g., Ref. \onlinecite{cano2014} for a derivation.} 
The $m^2 + n^2$ quasiparticles can be taken to be $\vec{l}_a = (a, 0)$, for $a = 0 , \cdots, m^2 + n^2 - 1$. 
The mutual braiding statistics between anyons labelled by $\vec{l}_a$ and $\vec{l}_b$ is 
therefore $e^{\frac{2 \pi i a b m }{m^2 + n^2}}$. 

Since we are interested in bosonic theories, we require $m$ to be even. 
In order for $\text{gcd}(m,n) = 1$, we require that $n$ be odd. This implies that
that $m^2 + n^2$ is odd. This corresponds to the anyon theory $\mathbb{Z}_N^{(p)}$ for $N = m^2 + n^2$ and $p = m/2$. 

When gauging charge conjugation in this theory, the resulting theory has $(N+7)/2$ anyons.\cite{dijkgraaf1989,barkeshli2014SDG} The anyons $[a], [-a]$ 
for $a = 1, \cdots, (N- 1)/2$ are grouped into non-Abelian anyons with quantum dimension 2. The identity particle
splits into two particles, $1, \epsilon$, with $\epsilon$ being the $\mathbb{Z}_2$ charge. After gauging
there are two types of twist fields, $\sigma_{\pm}$. These have fusion rules
\begin{align}
\sigma_+ \times \sigma_+ &= 1 + \sum_a \phi_a ,
\nonumber \\
\sigma_+ \times \sigma_- &= \epsilon +  \sum_a \phi_a .
\end{align}

Let us compute $\theta_\sigma^2$. As shown in Ref. \onlinecite{barkeshli2014SDG}, we have
\begin{equation}
	\theta_\sigma^2=\frac{\kappa_\sigma}{\sqrt{N}}\sum_{j=0}^{N-1}e^{\frac{2\pi i}{N}\frac{p(N-1)}{2}j^2}\equiv \kappa_\sigma G(N, p),
	\label{}
\end{equation}
where\cite{berndt1998}
\begin{equation}
	G(N,p)= \left( \frac{p(N-1)/2}{N} \right)G(N,1),
	\label{}
\end{equation}
$\left( \frac{a}{b} \right)$ is the Jacobi symbol, and
\begin{equation}
	G(N,1)= 
	\begin{cases}
		1 & N\equiv 1\,(\text{mod }4)\\
		i & N\equiv 3\,(\text{mod }4)
	\end{cases}.
	\label{}
\end{equation}
$\kappa_\sigma = \pm 1$ is the Frobenius-Schur indicator of $\sigma$, which depends on the choice of
$\mathcal{H}^3(\mathbb{Z}_2, \mr{U}(1)) = \mathbb{Z}_2$.\cite{barkeshli2014SDG} 
Therefore, for $N\equiv 1\,(\text{mod }4)$, we find $\theta_\sigma^2=\pm 1$.

Depending on the choice of $\mathcal{H}^3(\mathbb{Z}_2, \mr{U}(1)) = \mathbb{Z}_2$ class for the gauging process,
we have $\theta_{\sigma_{\pm}} = \pm 1$ or $\theta_{\sigma_{\pm}} = \pm i$. It is clear that the latter case gives 
rise to the inconsistency in $\eta_{\epsilon}^{\bf T}$, as discussed in Sec. \ref{conflictingEta}. The latter
case also implies that
\begin{align}
\mathcal{Z}(\mathbb{RP}^4) = \frac{1}{2{ N}}(1 + \eta_\epsilon^{\bf T}) \neq \pm 1. 
\end{align}
Thus gauging ${\bf T}^2$ gives us a theory with an $\mathcal{H}^3$ anomaly. As
discussed in Sec. \ref{Z4T}, this can be resolved by taking the true symmetry to be $\mathbb{Z}_4^{\bf T}$.
As in the examples of Sec. \ref{UspInfSeries}, we can now continue to gauge ${\bf T}^2$ again, and in this way
generate an infinite series of theories with $\mathcal{H}^3$ anomalies. 

In the more general case where $\text{gcd}(m,n) = f$, the fusion rules of the theory split as 
$\mathbb{Z}_{N/f} \times \mathbb{Z}_f$, with $N = m^2 + n^2$.\cite{cano2014} We leave a detailed analysis of this
case for future work. 

\section{Discussion}
\label{disc}

We have demonstrated a series of TQFTs which possess a $\mathbb{Z}_2^{\bf T}$ time-reversal 
symmetry localization anomaly, which is classified by $\mathcal{H}_{[\rho]}^3(\mathbb{Z}_2^{\bf T}, \mathcal{A})$. 
As described in Ref. \onlinecite{barkeshli2014SDG}, the general diagnostic of the existence of 
an $\mathcal{H}_{[\rho]}^3(\mathbb{Z}_2^{\bf T}, \mathcal{A})$ requires the full $F$ and $R$ symbols of the theory 
and their transformation properties under time-reversal. We have further provided a series of simpler constraints,
which only depend on the modular data (the topological spins and modular $S$-matrix), that must be satisfied 
for any (2+1)D theory to be free of this $\mathbb{Z}_2^{\bf T}$ localization anomaly. All of the theories that we considered
violated these constraints, signalling the existence of their $\mathcal{H}^3$ anomaly. Analogous 
results hold for $\mathbb{Z}_2^{\bf R}$ reflection symmetry, which we obtain by replacing the action of 
${\bf T}$ with ${\bf C R}$. 

There are a number of interesting questions that we leave for future work. In particular, it is unclear
whether theories with $\mathbb{Z}_2^{\bf T}$ symmetry localization anomalies always fall within the framework
found in this paper. For example, do all theories with $\mathcal{H}^3_{[\rho]}(\mathbb{Z}_2^{\bf T},\mathcal{A})$ anomalies
always violate the constraints that we have found and, if so, are any of the constraints always
violated together? In the examples that we have studied many of the constraints that we have found
are violated simultaneously. Furthermore, in our examples, in addition to the example of Ref. \onlinecite{fidkowski2015}
which studied a theory with unitary $\mathbb{Z}_2$ symmetry, the anomalous theories can be ``pseudo-realized'' at the
surface of a (3+1)D dynamical $\mathbb{Z}_2$ gauge theory with bosonic gauge charge; it would be interesting
to know whether cases with fermionic gauge charge exist as well. 

We note that the existence of a subgroup of Abelian anyons $\mathcal{A}$ implies the existence of a 
one-form symmetry with symmetry group $\mathcal{A}$.\cite{gaiotto2014} The cohomology class 
$\mathcal{H}^3_{[\rho]}(G, \mathcal{A})$ also appears in the study of $2$-groups, where $G$ is interpreted as the 0-form
symmetry, and $\mathcal{A}$ is interpreted as the 1-form symmetry.\cite{kapustin2013} This suggests another possible interpretation
of the $\mathcal{H}^3_{[\rho]}(G, \mathcal{A})$ anomaly as implying that the true symmetry of the TQFT is a non-trivial 
2-group symmetry, where the existence of the $\mathcal{H}^3_{[\rho]}(G, \mathcal{A})$ class implies that the 
$1$-form symmetry cannot be broken without also breaking the $0$-form symmetry. This 2-group symmetry then could
potentially possess a t' Hooft anomaly, which would be cancelled by a 2-group SPT in (3+1)D.\cite{kapustin2013,thorngren2015} 

However, an important point regarding the 2-group symmetry perspective is that 1-form symmetries apparently cannot be exact symmetries
of a system whose ultraviolet (UV) degrees of freedom are described by a Hilbert space that decomposes into a tensor product of local Hilbert 
spaces. Such models always contain dynamical matter fields in their effective quantum field theory description, which spoils the 
1-form symmetry at high enough energies. Therefore it appears that the 2-group symmetry perspective cannot resolve the possibility of whether the
symmetric topological phases of interest can exist in a system whose UV degrees of freedom form a Hilbert space that decomposes into a
tensor product of local Hilbert spaces. 

Furthermore, we note that Ref. \onlinecite{thorngren2015} studied t 'Hooft anomalies associated with 2-group symmetries
with non-trivial $\mathcal{H}^3$ class. There, the possibility of such anomalies being cancelled by a bulk (3+1)D 2-group SPT
via a generalization of anomaly in-flow was studied. Here we emphasize that the $\mathcal{H}^3$ anomalies described 
in Ref. \onlinecite{thorngren2015} are distinct from the $\mathcal{H}^3_{[\rho]}(G, \mathcal{A})$ discussed in this paper 
and those of Ref. \onlinecite{barkeshli2014SDG}. The $\mathcal{H}^3_{[\rho]}(G, \mathcal{A})$ anomalies
that we discuss here are defined solely by the action $[\rho]$, defined in Ref. \onlinecite{barkeshli2014SDG} and 
reviewed in Sec. \ref{H3review} of this paper. In particular, all Abelian topological phases where the symmetries 
do not permute the anyon types are free of such $\mathcal{H}^3_{[\rho]}(G, \mathcal{A})$ anomalies. 
The examples studied in Ref. \onlinecite{kapustin2014b,thorngren2015} uncover a different $\mathcal{H}^3$ structure in a 
purely Abelian topological phase with symmetries that do not permute the anyon types. There the obstruction arises once 
certain choices for the symmetry fractionalization class are made for a subset of the anyons, and one attempts to lift that 
choice to a symmetry fractionalization class for the full theory. It would be interesting to understand whether and how the 2-group anomaly in-flow picture 
also applies for  $\mathcal{H}^3_{[\rho]}(G, \mathcal{A})$ obstructions studied in this paper and in Ref. \onlinecite{barkeshli2014SDG}.

\section*{Acknowledgments}

We thank F. Benini, C. Cordova, P.-S. Hsin, A. Kapustin, N. Seiberg, R. Thorngren, K. Walker, and E. Witten
for discussions. We are especially grateful to N. Seiberg for bringing to our attention the time-reversal invariant
theories of Ref. \onlinecite{aharony2016} and for extensive discussions regarding the 
$\mathcal{H}^3$ anomaly. MC would like to thank Y. Qi for discussions on related topics. MB is supported by startup funds from the 
University of Maryland and NSF-JQI-PFC. We thank the Simons center at Stony Brook for hosting the 
program ``Mathematics of Topological Phases,'' where part of this work was completed. 

\appendix 

\section{Explicit Construction of Abelian Theories with $\Z_4^\mb{R}$ Symmetry}
\label{coupledWireAppendix}

In Sec. \ref{KmatrixExSec} we presented a series of $\mr{U}(1) \times \mr{U}(1)$ CS theories described by a $2\times 2$ K-matrix
$K = \left(\begin{matrix} m & n \\ n & -m \end{matrix} \right)$, which have a $\mathbb{Z}_4^{\bf R}$ 
reflection symmetry. Here we provide a construction of this theory that demonstrates that it possesses
no SPT (t' Hooft) anomaly. The idea is similar to ideas used in Ref. \onlinecite{song2016} and in Sec. \ref{pseudo2D}. 

We use the fact that the K-matrix theory can be realized as the effective theory of a system with a 
microscopic charge-conjugation symmetry. That is, $\mb{C}=-\mathds{1}$ is not only a symmetry of 
the effective topological field theory, but also corresponds to a microscopic on-site (internal) $\mathbb{Z}_2$ symmetry 
(which will also be denoted by $\mb{C}$ for convenience) of some lattice realization of the phase. 
This follows from the fact that in this theories, $\mb{C}$ is free of both the $\mathcal{H}^3_{[\rho]}(G,\mathcal{A})$ and $\mathcal{H}^4(G, \mr{U}(1))$
anomalies.\cite{barkeshli2014SDG}

We imagine cutting the two-dimensional space of the system into two regions, left and right, which are mapped to each other
under $\mb{R}$. By defining an appropriate action of $\mb{R}$, we show that the edge modes along the interface can be gapped
in a way which preserves the symmetry, without needing any additional degrees of freedom on the mirror plane of any bulk (3+1)D 
system. This will show that the action of $\mb{R}$ that we define is free of any SPT (t 'Hooft) anomalies. 

We define the $\mb{R}$ symmetry by composing the usual reflection, which only acts on the spatial degrees of freedom, 
with $\mb{C}$ restricted to the right half of the system (this is well-defined because $\mb{C}$ is on-site). On the edge fields, we have
\begin{equation}
	\mb{R}:\Phi_l \rightarrow -\Phi_r, \Phi_r \rightarrow \Phi_l.
	\label{eqn:z4r}
\end{equation}
Such a $\mb{R}$ transformation generates a $\mathbb{Z}_4^\mb{R}$ symmetry group.

Using the $\mb{C}$ symmetry, the entire K matrix of the interface edge theory is given by
\begin{equation}
	\mathcal{K}=
	\begin{pmatrix}
		K & 0\\
		0 & K\\
	\end{pmatrix}.
	\label{}
\end{equation}

We define the following two null vectors:
\begin{align}
	\bm{\Lambda}_1 &=\frac{1}{f}(m,n,-n,m), 
\nonumber \\
\bm{\Lambda}_2 & =\frac{1}{f}(n,-m,m,n).
\end{align}
Here $f=\text{gcd}(m,n)$.
We then add the following gapping terms to the Lagrangian:
\begin{equation}
	\begin{split}
	\delta \mathcal{L}&=-u ( \cos \bm{\Lambda}_1^\mathsf{T}K\Phi + \cos \bm{\Lambda}_2^\mathsf{T}K\Phi)\\
	&=-u\Big[\cos \frac{m^2+n^2}{f}(\phi_{l1}-\phi_{r2}) + \cos \frac{m^2+n^2}{f}(\phi_{l2}+\phi_{r1})\Big]
	\end{split}
	\label{}
\end{equation}
It is easy to see that $\delta \mathcal{L}$ preserves the reflection symmetry defined in Eq. \eqref{eqn:z4r}, and leads to a unique ground state. The cosine potentials pin $\phi_{l1}=\phi_{r2}, \phi_{l2}=-\phi_{r1}$ in the ground state, and therefore the two regions are joint together, such that the quasiparticles can tunnel between the two regions. In other words, we have realized a single topological phase described by the K matrix, with a $\mathbb{Z}_4^\mb{R}$ symmetry.

\bibliography{TI}

\end{document}